\documentclass[aps,prb,reprint,showpacs,showkeys,superscriptaddress]{revtex4-1}
\usepackage{multirow}
\usepackage{graphicx}
\usepackage{bm}
\usepackage{dcolumn}
\usepackage{picture}
\usepackage[hidelinks]{hyperref}
\usepackage{natbib}
\usepackage{amsmath}
\usepackage{times}
\usepackage{color}
\usepackage{soul}
\begin{document}
	
	\title{Magnetic disorder and gap symmetry in optimally electron doped Sr(Fe, Co)$_2$As$_2$ superconductor}
	
	\author{Luminita Harnagea} 
	\email[email:]{luminita@iiserpune.ac.in}
	\affiliation{Department of Physics, 
	Indian Institute of Science Education and Research, Pune, Maharashtra-411008, India}
	
	\author{Giri Mani}\affiliation{Department of Physics, Indian Institute of Science Education and Research, Pune, Maharashtra-411008, India}
	
	\author{Rohit Kumar}\affiliation{Department of Physics, Indian Institute of Science Education and Research, Pune, Maharashtra-411008, India}
	
	\author{Surjeet Singh}
	\email[email:]{surjeet.singh@iiserpune.ac.in}
%	\homepage[url:]{http://www.iiserpune.ac.in/~surjeet.singh}
	\affiliation{Department of Physics, Indian Institute of Science Education and Research, Pune, Maharashtra-411008, India}\affiliation{Center for Energy Science, Indian Institute of Science Education and Research, Pune, Maharashtra-411008, India}
	
	\date{\today}

	\begin{abstract}
	We investigate the magnetic pair-breaking due to Mn impurities in the optimally electron doped Sr(Fe$ _{0.88} $Co$ _{0.12} $)$ _{2} $As$_2 $ superconductor to deduce the symmetry of the superconducting order parameter. Experiments on the as-grown crystals reveled a T$ _c $ suppression rate of $ \sim $30 mK/$\mu \Omega cm$, which is in close agreement with similarly slower values of T$ _c $ suppression rates reported previously for various transition metal impurities, both, magnetic and non-magnetic, in several structurally analogous iron-based superconductors. However, careful annealing of these crystals at low temperature for longer durations reveals new information crucial to the determination of the pairing symmetry. We found that the crystallographic defects are a significant source of pair-breaking in the as-grown crystals. We first establish that these defects are point-like by showing that their sole effect on electrical transport is to add a temperature independent scattering term that shifts the whole $ \rho $ vs. T curves rigidly up. The T$ _c $ suppression rate due to these point-like defects is slow, $ \le$ 35 mK/$\mu \Omega cm$. On the other hand, T$ _c $ suppression rate due to magnetic pair-breaking is estimated to be faster than 325 mK/$\mu \Omega cm$. A slower pair-breaking rate (measured in mK/$\mu \Omega cm$) than expected due to non-magnetic crystallographic defects, together with a faster pair-breaking rate due to magnetic impurities disfavors a sign-changing $s_{+-}$-wave and argues in the favor of a non-sign-changing $s_{++}$-wave state in the optimally electron doped SrFe$ _2 $As$ _2 $ superconductor. 
	\end{abstract}


\maketitle

\section{Introduction}
The iron-based superconductors (FeSCs) have gathered considerable attention because of their high superconducting transition temperature(s) (T$ _c $) and intriguing superconducting properties (see Refs. \citenum{Hosono2015, Chubukov2015,Stewart2011, Johnston2010, Paglione2010, Ishida2009}). Since their discovered about a decade back, a considerable progress has been made towards understanding the normal state and superconducting properties of these materials\cite{Hosono2015, Chubukov2015}. However, the central question concerning the superconducting order parameter pairing symmetry has remained contentious till date \cite{Bang2017, Korshunov2017, Li2016rev, Hirschfeld2011}.Unlike the copper-based high-T$ _c $ SCs (called cuprates), where the d-wave symmetry of the order parameter was unequivocally established within few years of their discovery, no such universal pairing symmetry has been assigned to FeSCs. This is partly because of their complex electronic structure that generally consists of two or more hole-like Fermi sheets around the $\Gamma = (0, 0)$ point, and two electron-like Fermi sheets around the $M = (\pi, \pi)$ point of the 2-Fe Brillouin zone. 

In literature, one finds various competing theoretical proposals concerning the pairing symmetry in FeSCs (for a recent review on this, see Ref. \citenum{Hirschfeld2016}). However, a cast majority of these studies favor a fully gapped s-wave state in FeSCs. This s-wave state can either be a sign-changing $s_{+-}$ state where antiferromagnetic (AFM) spin-fluctuations are involved in the Cooper-pair formation \cite{Zhang2009,Mazin2008,Chubukov2008,Cvetkovic2009,Kempler2010,Hirschfeld2011} or a non-sign-changing s$ _{++} $ state where the orbital fluctuations are important for the pairing mechanism\cite{Kontani2010, Onari2012}. While both these states have the same symmetry, in the s$_{+-}$ case the superconducting order parameter changes sign between the hole and electron Fermi sheets, and in the s$_{++}$ model it preserves the sign. The fully gapped s-wave nature of the superconducting state has also been endorsed by several experiments, including penetration depth \cite{Malone2009, Hashimoto2009, Gordon2009, Gordon2009b, Kim2010c}, NMR \cite{Kawabata2008}, angle-resolved photoemission spectroscopy (ARPES) \cite{Ding2008} and $\mu$SR \cite{Luetkens2008}. However, these experiments cannot unambiguously tell whether the gap structure is $s_{+-}$ or s$_{++}$, which has remained a point of constant on-going debate \cite{Bang2017}.  

It has been suggested that pair-breaking in the presence of impurities can be a useful way to get around this problem. In a single-band s-wave superconductor, for instance, in the presence of a magnetic impurity the superconducting state is suppressed according to the Abrikosov Gor’kov (AG) law, which gives the T$ _c$ suppression rate as: $ ln(T_c/T_{c0}) = \psi(1/2) - \psi(1/2 + \gamma/2) $, where $T_{c0}$ is the superconducting transition temperature in the pristine sample, $ \psi $ is the digamma function and $ \gamma = \Gamma/ \pi T_c $ (see Ref. \citenum{Balatsky2006}). Here, $ \Gamma $ is the effective magnetic pair-breaking rate which is proportional to the impurity concentration. To account for T$ _c $ suppression in FeSCs, this formalism has also been extended to the isotropic multi-band s-wave superconductors \cite{Efremov2011}.  However, the experimentally observed T$ _c $ suppression rates in FeSCs, in the presence of various transition metal impurities, have been found to be almost an order or so in magnitude slower than what one would expect based on theory. To resolve this issue, Wang et al. (Ref. \citenum{Wang2013}) argued that a naive comparison of experimentally measured T$ _c $-suppression rates with theory in terms of impurity concentration is misleading. They proposed that a more useful way to compare T$ _c $-suppression rates is to express them in terms of increase in the residual resistivity upon impurity doping. Moreover, for a comparison with theory, the disorder should induce point-like scattering centers, which is not the case with chemically doped impurities that are not only non-point-like but can also potentially change the electronic structure by doping charge carriers in the system.

Recently, Prozorov et al. showed that in the superconductor Ba(Fe$_{1-x}$Ru$_x$)$ _2 $As$ _2 $ (x = 0.24), consisting mainly of electron irradiated point-like defects, a T$ _c $-suppression rate of about 350 mK/$\mu \Omega cm$ can be achieved\cite{Prozorov2014}. This rate is almost an order of magnitude faster than what was previously reported with chemically doped impurities, and can be described within the AG framework by considering the s$ _{+-} $ wave model. However, in Ba(Fe$_{1-x}$Ru$_x$)$ _2 $As$ _2 $, where replacement of Fe by Ru is an isovalent substitution, both types of charge carriers (i.e., electrons and holes) remain equally dominant up to high doping concentrations \cite{Rullier}; here, we ask, will the pairing symmetry still be s$ _{+-} $ if we take an optimally electron doped member whose Fermi surface will have the hole-pockets near the center of the Brillouin zone substantially shrunk, and the electron pockets at the zone corners enlarged \cite{Vilmercati}? Interestingly, we found evidence that in the optimally electron doped SrFe{$ _2 $}As{$ _2 $}, the gap structure conforms to the s$ _{++} $-wave state.   

The rest of the manuscript has been organized as follows. Section \ref{Experimental procedure} deals with the experimental details, including, the single-crystal growth and the annealing treatment. In section \ref{Composition and structural analysis} structural characterizations are presented. Thermopower, Hall effect and magnetic measurements are presented in section \ref{Spin and charge state of doped Mn impurities}. Section \ref{T$ _c $ suppression upon Mn doping} deals with investigation of T$_c$-suppression in the as-grown crystal, and the effect of annealing on the T$ _c $-suppression is presented under section \ref{Effect of gentle annealing on T$ _c $ suppression}. A discussion of the gap symmetry and pair-breaking mechanism based on results presented in the preceding sections is covered under section \ref{Discussion}. This is followed by section \ref{Summary and conclusions} which gives a summary of the important results and conclusion derived from this work.         

\section{Experimental procedure} \label{Experimental procedure}
Experiments were carried out on  a series of self-flux grown single crystals of compositions
Sr(Fe$ _{1 -x{_N} - y{_N}}$Co$_{x{_N}} $Mn$_{y{_N}}$)$ _2 $As$ _2$ (for $x_N = 0.14$ and $y_N = 0$ to $0.15$ (see table 1 for values of  $y_N$); here and elsewhere in the manuscript the symbols $ x_N $ and $ y_N $ are used to refer to the nominal compositions, and x and y to their corresponding experimentally determined values. The weighing, mixing and grinding of the precursors were carried out in an Ar-filled glove box where the level of moisture and O$_2$ is always maintained below 0.1 ppm. The growth experiments were carried out in two steps. In the first step, precursor materials, namely, FeAs, Co$_2$As and MnAs were prepared using the method similar to that described in ref. \citenum{Harnagea2011}. Briefly, stoichiometric quantities of high-purity Fe (Alpha Aesar, 99.90\%), Co (Alpha Aesar, 99.8\%) or Mn (Alpha Aesar, 99.95\%) powders were thoroughly ground and mixed with As powder (Alpha Aesar, 99.999\%) and filled in alumina crucibles which were loaded in pre-heated quartz ampules inside the glove-box. Immediately after taking them out of the glove-box, the ampules were sealed under high-vacuum condition and were subsequently heated to a temperature of 700$^{\circ}$C. They were maintained at this temperature for 10 hours before cooling to room-temperature. At the end of this reaction, the ampules were transferred inside the glove-box where they were cut open to remove the reacted products. These products were grounded and stored inside the glove-box.  

\begin{figure}
	\vspace{0.1cm}
	\hspace{0.5cm}
	\begin{center}
\includegraphics[scale=0.34]{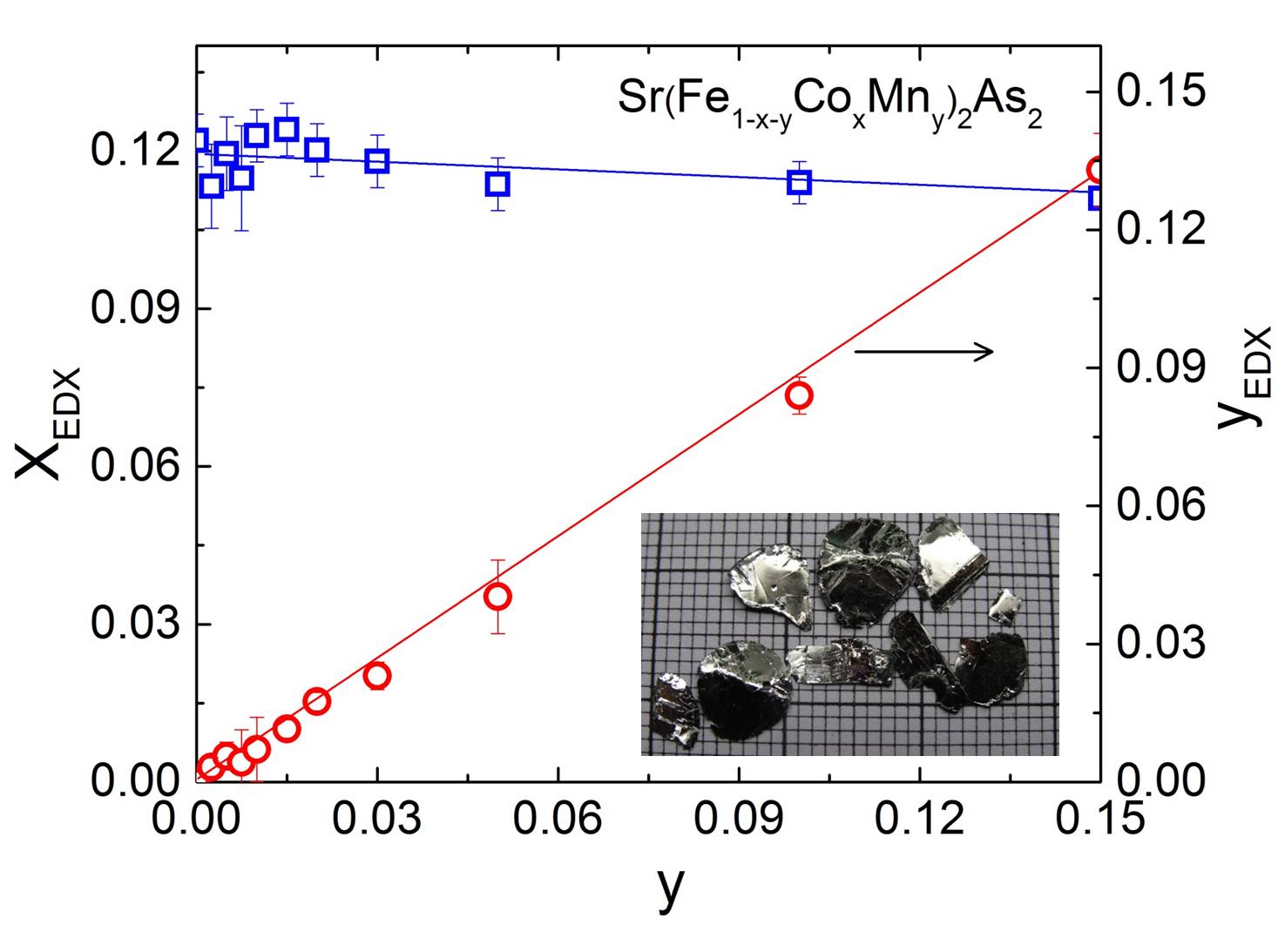}
\end{center}
\caption{Co and Mn concentrations determined using the EDX probe, and designated here as x$_{edx}$ and y$_{edx}$, respectively, are plotted against the nominal Mn composition ($y$). Inset shows a few representative single crystals of SrFe$_{2}$As$_2$.}
\label{FIG.1}
\end{figure}

In the second step, these metal-arsenic precursors were mixed with Sr metal pieces (Sigma Aldrich, 99.00\%). Appropriate quantities of As powder was added so to have the desired stoichiometry of Sr : Fe$_{(1-x{_N}-y{_N})}$Co$_{x{_N}}$Mn$_{y{_N}}$As$_2$ $\equiv$ 1: 4. The excess metal-arsenic is used as a self-flux. The growth experiments were carried out in alumina crucibles sealed under vacuum in quartz ampules. The charge was slowly heated to 1100$^{\circ}$C, kept there for 24 hrs and thereafter cooled slowly down to 950$^{\circ}$C at a rate of 3.5$^{\circ}$C per hour.  After waiting here for 1 hr, the temperature was decreased to ambient at a rate of 300$^{\circ}$C per hour. This growth procedure allowed us to obtain solidified \textit{cm}-size ingots from which single-crystals having lateral dimension up to 10 mm and thicknesses up to 0.25 mm were mechanically extracted. Representative images of the grown crystals are shown in the inset of Fig.\ref{FIG.1}.

The lattice parameters were obtained using the X-ray powder diffraction technique. For this purpose, fine powders obtained by crushing small single crystal pieces were used. Measurements were done using a Bruker diffractometer (D8 advanced) equipped with Cu $K\alpha$- radiation. Chemical composition, growth behaviour and morphology of the grown crystals was assessed using a scanning electron microscope (Zeiss Sigma – FESEM) equipped with energy dispersive X-ray analysis (EDX) probe. Electrical resistivity was measured from T = 2 to 300 K on rectangular platelet-like single crystals, using the standard four-probe technique in a Physical Property Measurement System (PPMS) from Quantum Design, USA. Current (I) and voltage (V) contact were made by gluing gold wires on the sample surface (//  \textit{ab}-plane) using a silver epoxy. Magnetization measurements were carried out using a superconducting quantum interference device magnetometer (MPMS – XL7) from Quantum Design, USA. Zero field cooled (ZFC) and field cooled (FC) magnetic data were recorded in a temperature range of 2-20 K under an applied magnetic field of 20 Oe. The magnetization data of several samples was also recorded in a temperature interval of 2-300 K under a magnetic field of 10 kOe applied parallel to the crystallographic \textit{ab}-plane. The Seebeck coefficient and Hall measurements were carried out in a PPMS ( Quantum Design, USA) using the standard measurement probes. 

\section{ Results}

\subsection{Composition and structural analysis of Sr(Fe$_{(1-x{_N}-y{_N})}$Co$_{x{_N}}$Mn$_{y{_N}}$)$_{2}$As$_{2}$ ($x{_N} = 0.14, 0 \leq  y{_N} \leq 0.15$)} \label{Composition and structural analysis}
The Sr(Fe, Co, Mn)$_{2}$As$_{2}$ single crystals exhibit a layered structure with large terraces, nearly of micrometer size, terminate at sharp steps. They tend to cleave easily along the \textit{ab}-plane, and showed tendency towards exfoliation. In order to determine the chemical composition of our single crystals, a few pieces from each growth experiment were analyzed in detail under a SEM. On each specimen, the composition is determined using EDX over 15 to 20 spots. From these data, the average Co and Mn composition is calculated. The statistically averaged composition and standard deviations were used to plot data shown in Fig.\ref{FIG.1}, where the averaged EDX composition is plotted against the nominal composition. The standard deviation in each case is found to be less than 0.5 at. \%. This value is within the error bar of the EDX technique ($\sim$ 1 at. \%) and indicates a fairly good homogeneity of Co, Mn distribution within a single crystal, and several single crystals from a given growth experiment. The actual Co composition of our single crystals varied from $ x = x_{EDX}$ $\approx 0.11$ to $ 0.12 $ (Fig.\ref {FIG.1}), which is slightly smaller than the nominal value  of $ 0.14 $. In a previous report on Sr(Fe, Co)$_{2}$As$_{2}$ single crystals grown using the self-flux technique, the superconducting dome is reported to extend from $x \approx  0.07$ to $ 0.17 $, with a maximum value of $T_c$ around $x  \approx  0.117$ (Ref. \citenum{Hu2011}). Since the variation of $T_c$ near the dome maximum is typically marginal, a minor variation in the Co-concentration for our variously Mn-doped crystals can be regarded as practically insignificant, and we can safely assume that the present investigation reveals the effect of T$ _c $ suppression due to Mn-doping in the optimally electron doped Sr122 superconductor.

As shown in Fig.\ref{FIG.1}, the Mn concentration obtained using EDX vary linearly as a function of the nominal composition. From a linear fit (solid red line in the Fig.\ref{FIG.1} given by $y = y_{EDX} = (0.872\pm 0.013)y_{N} $) one can obtain the experimentally determined concentration for any given nominal composition. In the rest of manuscript, we will refer to our single crystals by their EDX compositions shown in Table.\ref{table-comp}. The determined chemical formula being Sr(Fe$_{0.88-y}$Co$_{0.12}$Mn$_y$)$_2$As$_2$.

\begin{table*}
	\caption{\textbf{Chemical formula, nominal ($ y_N $) and the obtained Mn composition using $y = (0.872\pm 0.013)y_{N}$. Actual Co concentration $ (x) $ in each sample is close to 0.12 (see text for details).}}
	\label{table-comp}
	\renewcommand{\arraystretch}{1.3}
	\begin{center}
		\begin{tabular} {p{6cm} p{2.5cm} p{2.5cm} p{2.5cm}}
			\hline
			\textbf{Chemical formula} & \textbf{$y_N $} & \textbf{$ y $} & \textbf{Short name}\\
			\hline
			SrFe$ _2 $As$ _2  $ & 0 & 0 & Sr122 \\
			%\hline
			Sr(Fe$ _{0.86} $Co$ _{0.14} $)$ _2 $As$ _2 $ & 0 & 0 & Mn0.0\\
			%\hline
			Sr(Fe$ _{0.8575} $Co$ _{0.14} $Mn$ _{0.0025} $)$ _2 $As$_2 $ & 0.0025 & 0.002 & Mn0.2\\
			%\hline
			Sr(Fe$ _{0.855} $Co$ _{0.14} $Mn$ _{0.005} $)$ _2 $As$_2 $ & 0.005 & 0.004 & Mn0.4\\
			%\hline
			Sr(Fe$ _{0.8525} $Co$ _{0.14} $Mn$ _{0.0075} $)$ _2 $As$_2 $ & 0.0075 & 0.006 & Mn0.6 \\
			%\hline
			Sr(Fe$ _{0.85} $Co$ _{0.14} $Mn$ _{0.01} $)$ _2 $As$_2 $ & 0.01 & 0.009 & Mn0.9\\
			%\hline
			Sr(Fe$ _{0.845} $Co$ _{0.14} $Mn$ _{0.015} $)$ _2 $As$_2 $ & 0.015 & 0.013 & Mn1.3\\
			%\hline
			Sr(Fe$ _{0.84} $Co$ _{0.14} $Mn$ _{0.02} $)$ _2 $As$_2 $ & 0.02 & 0.017 & Mn1.7\\
			%\hline
			Sr(Fe$ _{0.83} $Co$ _{0.14} $Mn$ _{0.03} $)$ _2 $As$ _2  $ & 0.03 & 0.026 & Mn2.6\\
			%\hline
			Sr(Fe$ _{0.81 }$Co$ _{0.14} $Mn$ _{0.05} $)$ _2 $As$_2 $ & 0.05 & 0.044 & Mn4.4\\
			%\hline
			Sr(Fe$ _{0.76} $Co$ _{0.14} $Mn$ _{0.10} $)$ _2 $As$_2 $ & 0.10 & 0.087 & Mn8.7\\
			%\hline
			Sr(Fe$ _{0.71} $Co$ _{0.14} $Mn$ _{0.15} $)$ _2 $As$_2 $ & 0.15 & 0.131 & Mn13\\ 
			\hline
					\end{tabular}
	\end{center}
\end{table*}

\begin{figure}[!]
	% \vspace{-0.1cm}
	% \hspace{2cm}
	\centering
	\includegraphics[height = 6 cm]{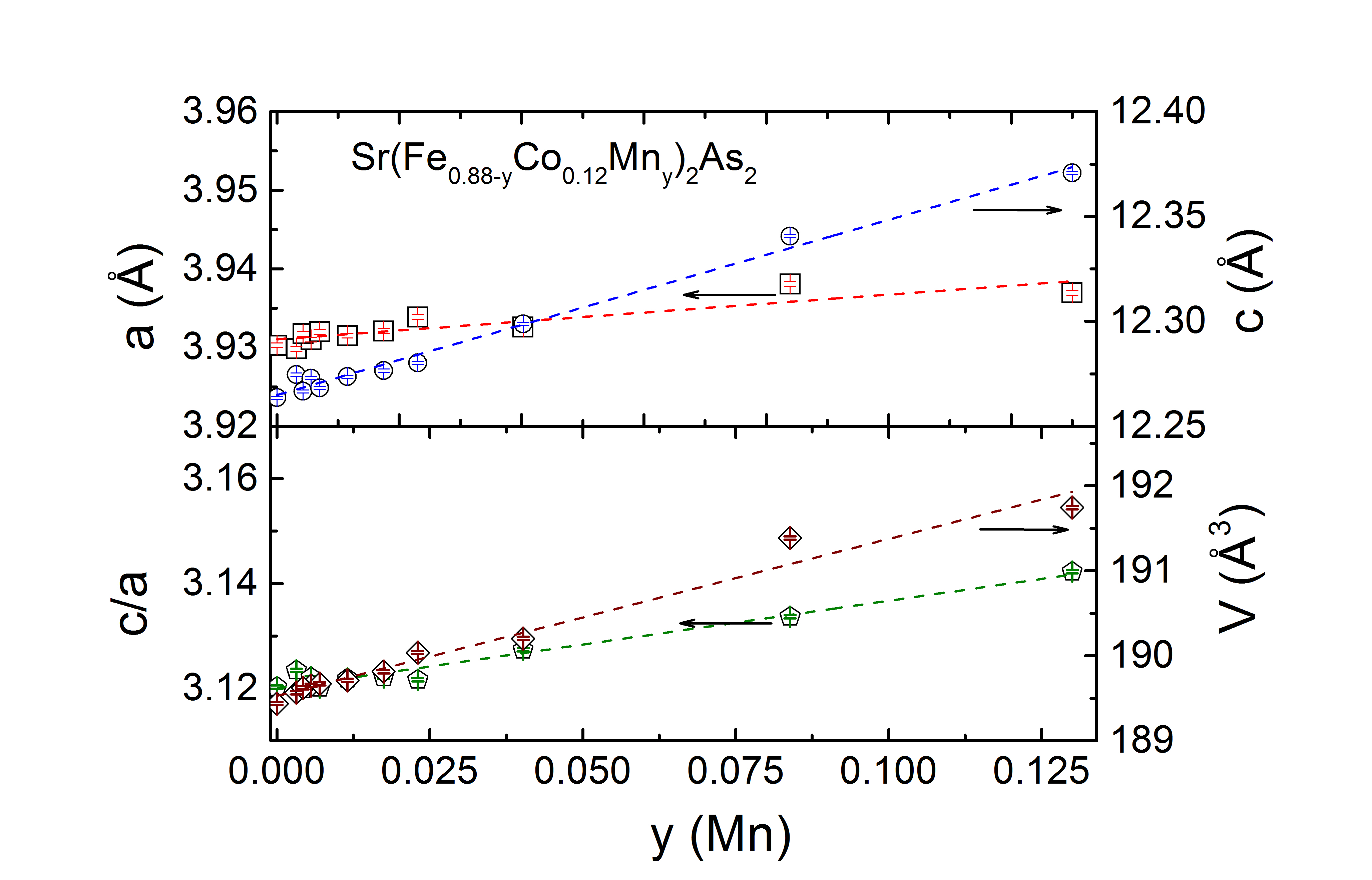}
	\caption{Lattice parameters \textit{a}, \textit{c}, \textit{c/a} and unit cell volume V of Sr(Fe$_{0.88-y}$Co$_{0.12}$Mn$_y$)$_2$As$_2$  single crystals plotted as a function of actual Mn concentration ($ y $) (see text for details)}
	\label{FIG.2}
\end{figure}
The phase purity of our single crystals and the lattice parameters were determined using x-ray powder diffraction. Several single crystal pieces from each batch were ground into a fine powder along with silicon, taken as an internal standard; and the diffraction pattern was recorded at room temperature. The samples were found to be single phase and only arbitrarily we observed additional diffractions peaks due to the residual flux, which might have remained on the surface of crystal prior to grinding. The observed powder patterns were indexed based on ThCr$_{2}$Si$_2$ tetragonal structure, space group I4/mmm, No. 139; and the lattice parameters were determined using the UNITCELL refinement software. Additionally, we checked the crystallinity and orientation of the grown crystals by performing x-ray diffraction on cleaved platelet-like crystals in the Bragg-Brentano geometry, which yielded  diffraction patterns showing only peaks with Miller indices (0, 0, 2l), indicating that the \textit{c}-axis is perpendicular to the plane of platelets.  This was also confirmed using  the back-reflection Laue diffraction technique. 

The lattice parameters, unit cells volume and \textit{c/a} variation across the Sr(Fe$_{0.88-y}$Co$_{0.12}$Mn$_y$)$_2$As$_2$ series as a function of Mn concentration are shown in the Fig.\ref{FIG.2}. The lattice parameters of SrFe$_{2}$As$_2$ are found to be $a_0$ = 3.928 {\AA} and $c_0$ = 12.354 {\AA}, in good agreement with previous reports \cite{Hu2011,Yan2008}. Substitution of Co ($x = 0.12$) for Fe in SrFe$_{2}$As$_2$ induces a minor increase in the value of \textit{a}  ($ a $ = 3.930 {\AA} ), while the \textit{c}-parameter showed a significant decrease ($ c $ = 12.264{\AA}). These variations are in agreement with the previous report \cite{Hu2011}. Upon substitution of Mn for Fe in Sr(Fe$_{0.88}$Co$_{0.12}$)$_2$As$_2$, the lattice parameters show a gradual increase (Fig. \ref{FIG.2}). For the crystal with the highest Mn concentration, we found $a/a_0 = 1.002$, $c/c_0 = 1.001$ and $V/V_0 = 1.006$ (where $a_0$, $c_0$, $V_0$ are lattice parameters of SrFe$_{2}$As$_2$). These trends are in agreement with previous reports on Mn-doped Sr122 and Ba122 series of compounds \cite{Kim2010,Thaler2011}; and are also consistent with the differences in the ionic radii, which for the four-fold coordinated $Fe^{2+}$, $Co^{2+}$ and $Mn^{2+}$  is 0.63 {\AA}, 0.58 {\AA} and 0.66 {\AA}, respectively. Since Mn-ion in higher oxidation states has an ionic radius less than 0.58 {\AA}, the observed increase in the unit cell volume indicates that doped Mn ions are in their +2 oxidation state. 

\subsection{Spin and charge state of doped Mn impurities} \label{Spin and charge state of doped Mn impurities}
\begin{figure}[!]
	% \vspace{-0.9cm}
	% \hspace{0.1cm}
	\begin{center}
		\includegraphics[scale=0.78]{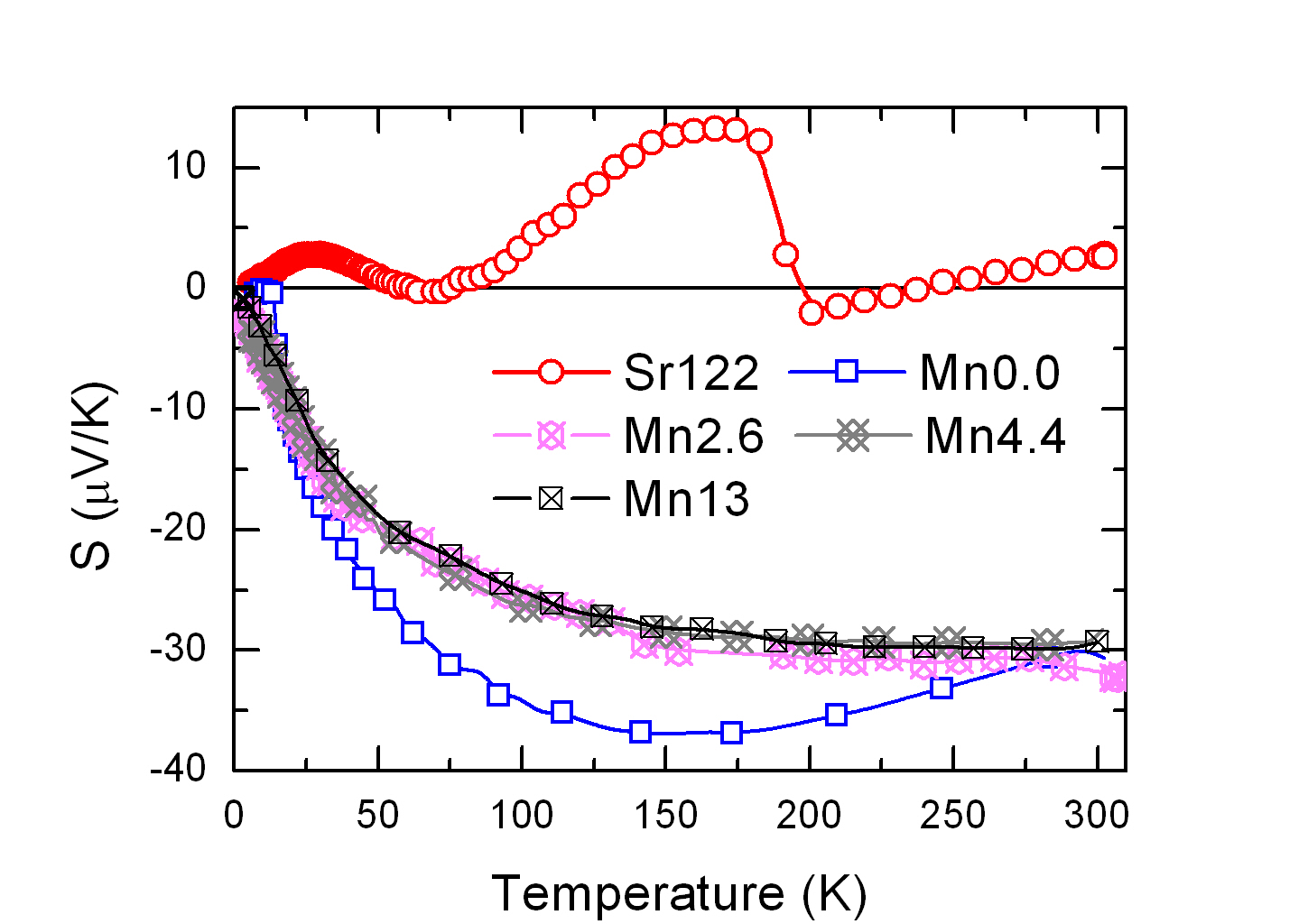}
	\end{center}
	\caption{In-plane Seebeck coefficient (S) of SrFe$_{2}$As$_2$ and Sr(Fe$_{0.88-y}$Co$_{0.12}$Mn$_y$)$_2$As$_2$ (y = 0, 0.026, 0.044, 0.13) single crystals plotted as a function of temperature.}
	\label{seeb}
\end{figure}

 The temperature (T) dependence of in-plane thermopower (S) of SrFe($ _{0.88-y} $Co$ _{0.12} $Mn$ _{y} $)$ _{2} $As$ _{2} $ for several representative samples, namely, Mn0.0, Mn2.6, Mn4.4, Mn13, and Sr122 is shown in Fig.\ref{seeb}. Thermopower of Sr122 near T = 300 K is small and positive ($ \sim 3 \mu$V/K ), and it gradually decreases upon cooling the sample and changes sign near T = 250 K. Upon further cooling, it changes sign again with a step-like increase to a relatively large and positive value of 12 $\mu$V/K near T = 190 K. This feature coincides with the combined structural/magnetic transition previously reported in Sr122, indicating a significant reconstruction of Fermi surfaces across this transition, in agreement with previous Hall and ARPES data\cite{Zhang2009,Chen_2008}. Below the transition, thermopower first decreases to zero (near 60 K) before increasing again resulting in a small hump around 25 K in good agreement with the previous reports \cite{Sasmal2008, Butch2010}. Overall, the thermopower behavior of Sr122 is complex, which appears to be a common feature of the parent compounds in FeSC families and reflects an interesting interplay of multiband electronic structure and electron-phonon interactions (phonon drag) in these materials (for a review see ref. \citenum{Pallecchi2016}).   
 
In the optimally Co-doped sample Sr(Fe$ _{0.88} $Co$ _{0.12} $)$ _{2} $As$ _{2} $ (Mn0.0), the thermopower behavior changes dramatically with respect to the undoped compound (Fig.\ref{seeb}). It now varies smoothly and remains negative over the whole temperature range except in the superconducting state where it becomes zero as expected. The negative sign of thermopower suggests that substituting Co for Fe in SrFe$ _{2} $As$ _{2} $ effectively results in electron doping in the material, in agreement with the ARPES results \cite{Zhang2009}. The thermopower attains its maximum negative value of 38 $\mu V/K$ around T = 150 K in good agreement with the optimally Co-doped sample in the  Ba(Fe$_{1-x}$$Co_{x}$)$_{2}$As$_{2}$ series \cite{Mun2010,Yan2010}.

 With the substitution of Mn, the thermopower remains negative and smoothly varying over the entire temperature range. It decreases slightly with initial Mn-doping. The reason for this decrease is not investigated further. It may be possible that the magnitude of thermopower in Sr(Fe$ _{0.88} $Co$ _{0.12} $)$ _{2} $As$ _{2} $ is sensitive to the crystal orientation in the \textit{ab}-plane. However, what is more interesting is that the thermopower remains unchanged for samples with different Mn concentrations up to 13 \% of doping level. This apparent insensitivity of the measured thermopower to an increasing Mn concentration suggests that Mn substitution at the Fe site does not alter the charge carrier concentration appreciably. The temperature variation of Hall coefficient of samples Mn0.0 and Mn2.6, shown in the supplementary figure 10, also supports this conclusion. 

The localized nature of doped Mn electrons is also inferred from the magnetization measurements shown in supplementary information (figure 12) for samples Sr122, Mn0.0 and Mn13. The temperature variation of magnetization of sample Sr122 is comparable with that reported  previously \cite{Hu2011,Yan2008}. Sample Mn0.0 is superconducting with a T$ _c $ onset of about 12 K. On the other hand, sample Mn13 is non-superconducting, and at low-temperatures (i.e., below about 50 K), its magnetization increases upon cooling in a Curie-like manner. A Curie-Weiss analysis of the low-T data reveals that the doped Mn-ions are in their $ +2 $  oxidation state carrying an effective spin S $ = 1/2 $, which corresponds to the low-spin state of Mn in a tetrahedral ligand coordination. This conclusion concerning the spin and charge state of doped Mn ions is consistent with previous nuclear magnetic resonance (NMR), electron spin resonance (ESR) and inelastic neutron scattering studies on Mn-doped  BaFe$_{2}$As$_{2}$ \cite{Texier2012,Rosa2014a,Tucker2012}.        

\subsection{T$ _c $ suppression upon Mn doping} \label{T$ _c $ suppression upon Mn doping}
\begin{figure}[t]
	%\vspace{.1cm}
	%\hspace{0.2cm}
	%\begin{center}
	\includegraphics[scale=0.78]{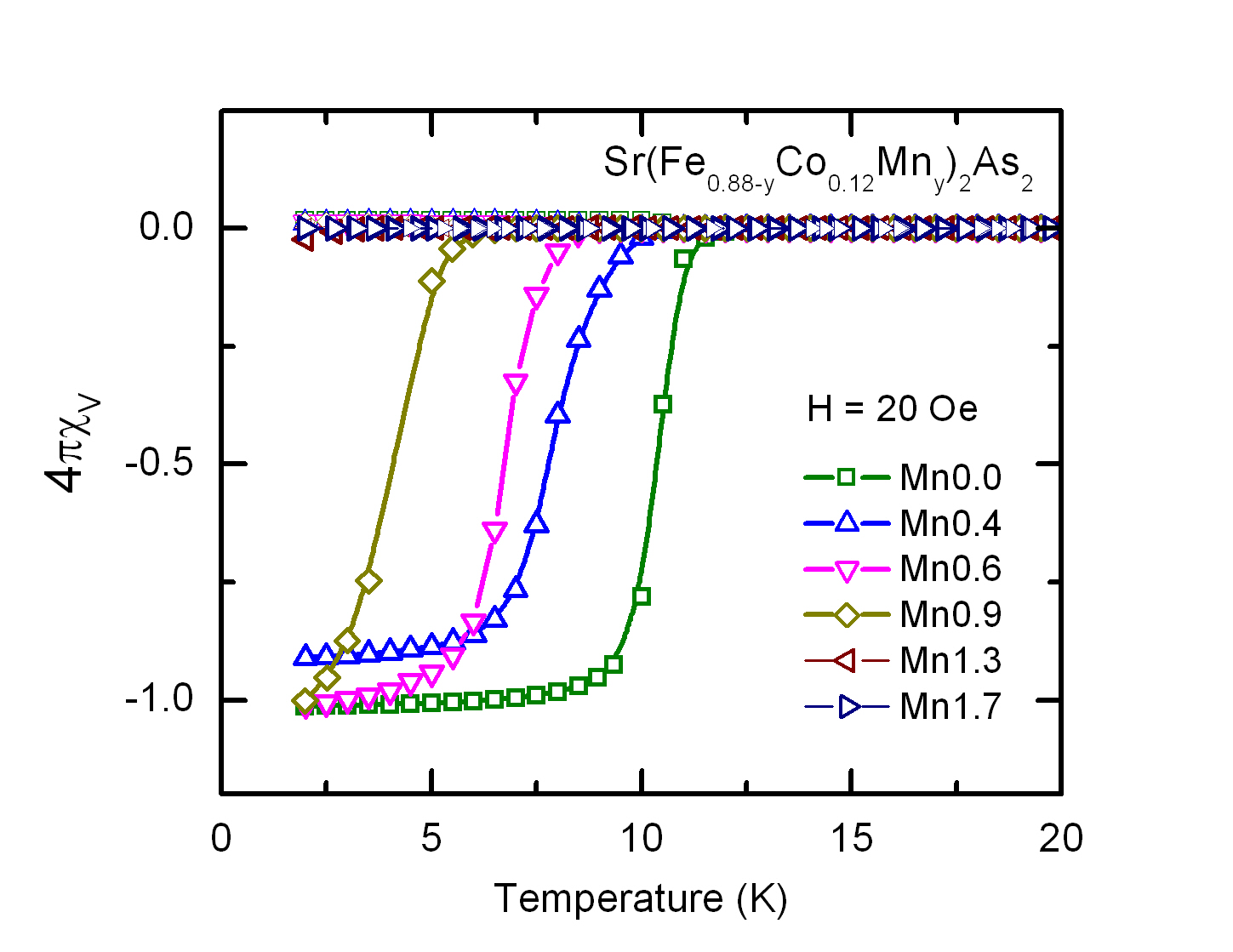}
	\caption{Temperature dependence of volume susceptibility ($ \chi_V $) plotted as $ 4\pi \chi_V $ for representative    Sr(Fe$_{0.88-y}$Co$_{0.12}$Mn$_y$)$_2$As$_2$ (y = 0, 0.004, 0.006, 0.009, 0.013 and 0.017) single crystals measured under a field (H) of 20 Oe applied parallel to the surface (\textit{ab}-plane) of the platelet shaped crystals.}
	\label{chi}
	%\end{center}
\end{figure}
The zero-field-cooled (ZFC) and field-cooled (FC) volume susceptibilities ($\chi_V$) of a few representative samples is shown in Fig. \ref{chi}. Measurements were done under an external magnetic field of 20 $ \pm $ 5 Oe applied parallel to the \textit{ab}-plane. Transition to a superconducting state upon cooling is evident from the large diamagnetic signals in the ZFC scans for samples Mn0.0, M0.4, Mn0.6 and Mn0.9. Sample Mn1.3 shows a superconducting onset near T = 3 K but the diamagnetic signal remained very small down to the lowest measurement temperature. No sign of superconductivity is observed in sample Mn1.7 down to T = 2 K.

The superconducting temperature (T$ _c(\chi) $) is obtained as the point of maximum rate change of $ \chi(T) $ with respect to temperature below the superconducting onset. Using this criterion, T$ _c$ values of 11, 9, 7.5 and 5 K are estimated for samples Mn0.0, Mn0.4, Mn0.6 and Mn0.9, respectively. T$ _c $ of the sample Mn0.0 in agreement with previous reports \cite{Hu2011}. For all the samples up to 0.9 \% of Mn doping, the superconducting volume fraction is close to $100 \%$ ($4\pi\chi_V \sim -1$). The corrections due to demagnetization factor using the ellipsoid approximation \cite{Osborn1945} is estimated to be less than 10 \% , which ensure bulk superconductivity in samples up to 0.9 \% of Mn doping. The unsystematic variation in teh value of $\chi_V$ in the superconducting state is probably due to the residual magnetic field in the magnetometer. For the sample M1.7 the superconducting volume fraction remains marginally small down to 2 K, which indicating that the critical Mn concentration required to quench the superconductivity completely in optimally electron doped Sr122 is close to this value.

\begin{figure}[!]
	%\vspace{.1cm}
	%\hspace{0.2cm}
	%\begin{center}
\includegraphics[scale=0.7]{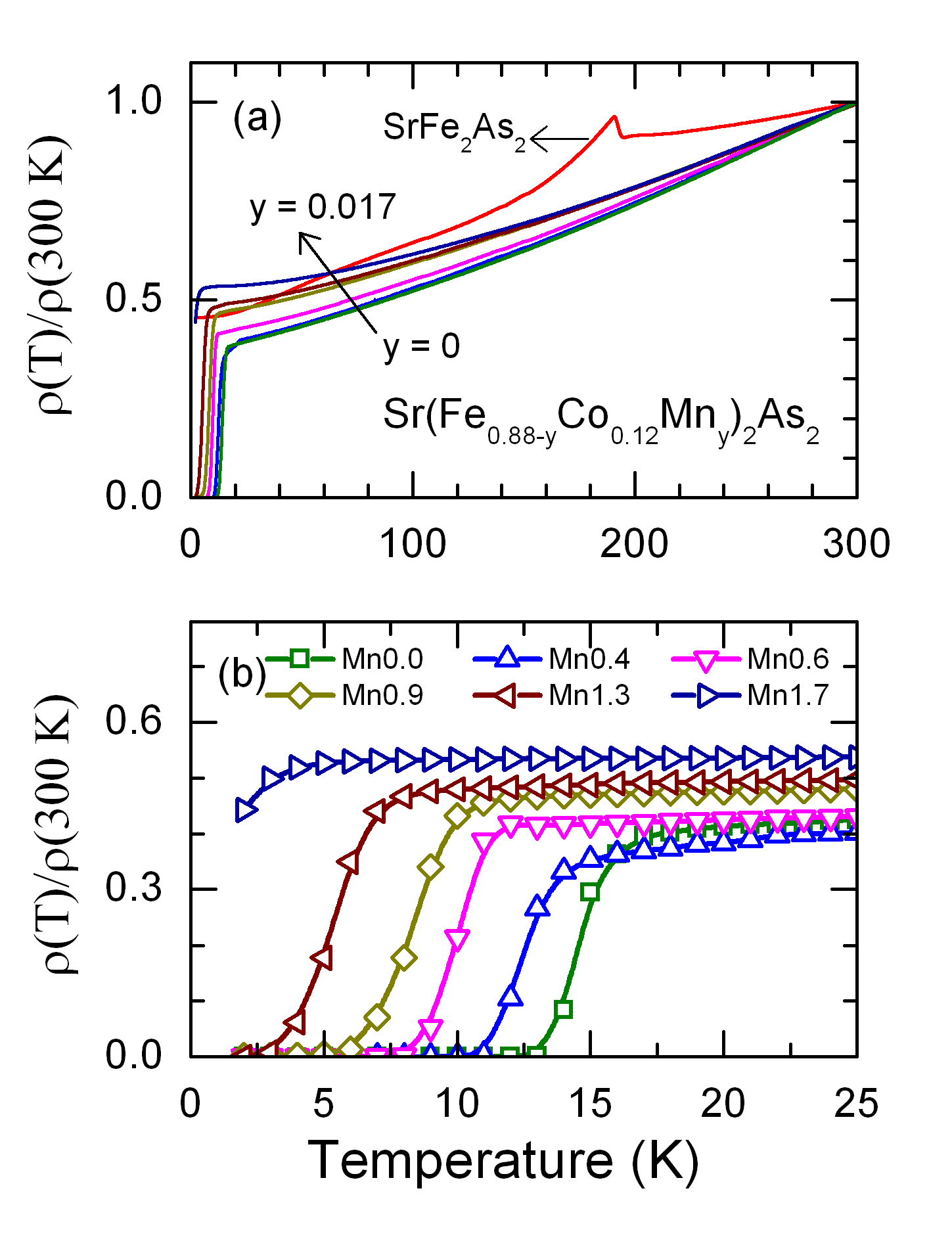}
\caption{Normalized resistivity R(T)/R(300 K) of the as-grown Sr(Fe$_{0.88-y}$Co$_{0.12}$Mn$_y$)$_2$As$_2$ single crystals. Lower panel is an enlarged view of the low-temperature region of panel (a) showing the superconducting transitions.}
\label{rhosc}
%\end{center}
\end{figure}

We further investigate the T$_c$ suppression using the electrical transport measurements. The temperature variation of normalized in-plane resistivity ($\rho^n(T)$ = R(T)/R(300 K)), where R(T) is the measured resistance at any temperature T and R(300 K) at T = 300 K, is shown in Fig. \ref{rhosc}. Sample SrFe$ _{2} $As$ _{2} $ exhibits a metallic behavior over the entire temperature range (Fig. \ref{rhosc}a), with a distinct anomaly near $\sim$ 192 K (T$ _o $). This anomaly corresponds to the simultaneous structural and magnetic phase transitions from a tetragonal-paramagnetic to an orthorhombic-antiferromagnetic phase upon cooling across T$ _o $ (Ref. \citenum{Yan2008}). The position and step-like appearance of this anomaly is in a good agreement with the previous reports \cite{Hu2011,Sefat2014}. In sample Mn0.0, the structural/magnetic transition is fully suppressed, which is now replaced by a superconducting transition at low-temperatures, as shown in an expanded view in panel (b) of Fig. \ref{rhosc}. The superconducting transition temperature is determined using the zero-resistance criteria (i.e., the temperature at which the resistance of sample first becomes zero upon cooling). The value of T$ _c $ for sample Mn0.0 is found to be 12.5 K, which is in good agreement with the T$ _c $ reported by Hu et al \cite{Hu2011} for their optimally Co-doped doped Sr122 single crystal grown using the self-flux technique. 

Doping with Mn at the Fe-site in Sr(Fe$_{0.88}$Co$_{0.12}$)$_2$As$_2$ results in a gradual suppression of T$ _c$ (Fig. \ref{rhosc}b). In crystals Mn0.4, Mn0.6, Mn0.9 and Mn1.3, $T_c$ estimated using the zero-resistance criteria has been found to have suppressed values: $ 11 $, $ 8.5  $, $ 5.5 $ and $ 2 $ K, respectively. These values are in close agreement with T$ _c $ obtained from the diamagnetic signal in the volume susceptibility. For sample Mn1.7, only a partial drop in resistivity below T = 3 K is observed, in agreement with the magnetization data. Superconducting transition width $\bigtriangleup T_c$ for these crystals is estimated using: $\bigtriangleup T_{c} $= T$_c^{onset}$ - T$_c^{zero}$. Here, T$_c^{onset}$ is the temperature below which the resistivity starts dropping from its normal-state value. The ratio $\bigtriangleup T_c/T _c^{mid}$, where T$_c^{mid}$ = (T$_c^{onset}$ + T$_c^{zero}$)/2, for Mn0.0 is about 0.27. This value is comparable to the values previously reported for other FeSCs (see Refs. [\citenum{Saha2009, Hu2011}] for $\bigtriangleup T_c/T _c$ values in Co and Ni-doped Sr122 crystals). This ratio increases with increasing Mn concentration reaching a value of 0.33 for sample Mn0.9. For Mn1.3, this value exceeds 1 due to spurious superconducting onset, as shown using the $\chi_V$ data in Fig. \ref{chi}. This broadening of the transition width correlates with the increasing degree of structural disorder associated with random site occupancy of Mn-ions replacing the Fe-ions in FeAs layers. Additionally, nanoscale inhomogeneities in the the Mn-concentration across the sample volume cannot be ruled out which also contribute to the broadening of superconducting transition. We shall return to this point again while discussing the results.  

\begin{figure}[]
	%\vspace{-0.1cm}
	%\hspace{0.1cm}
\includegraphics[scale=0.7]{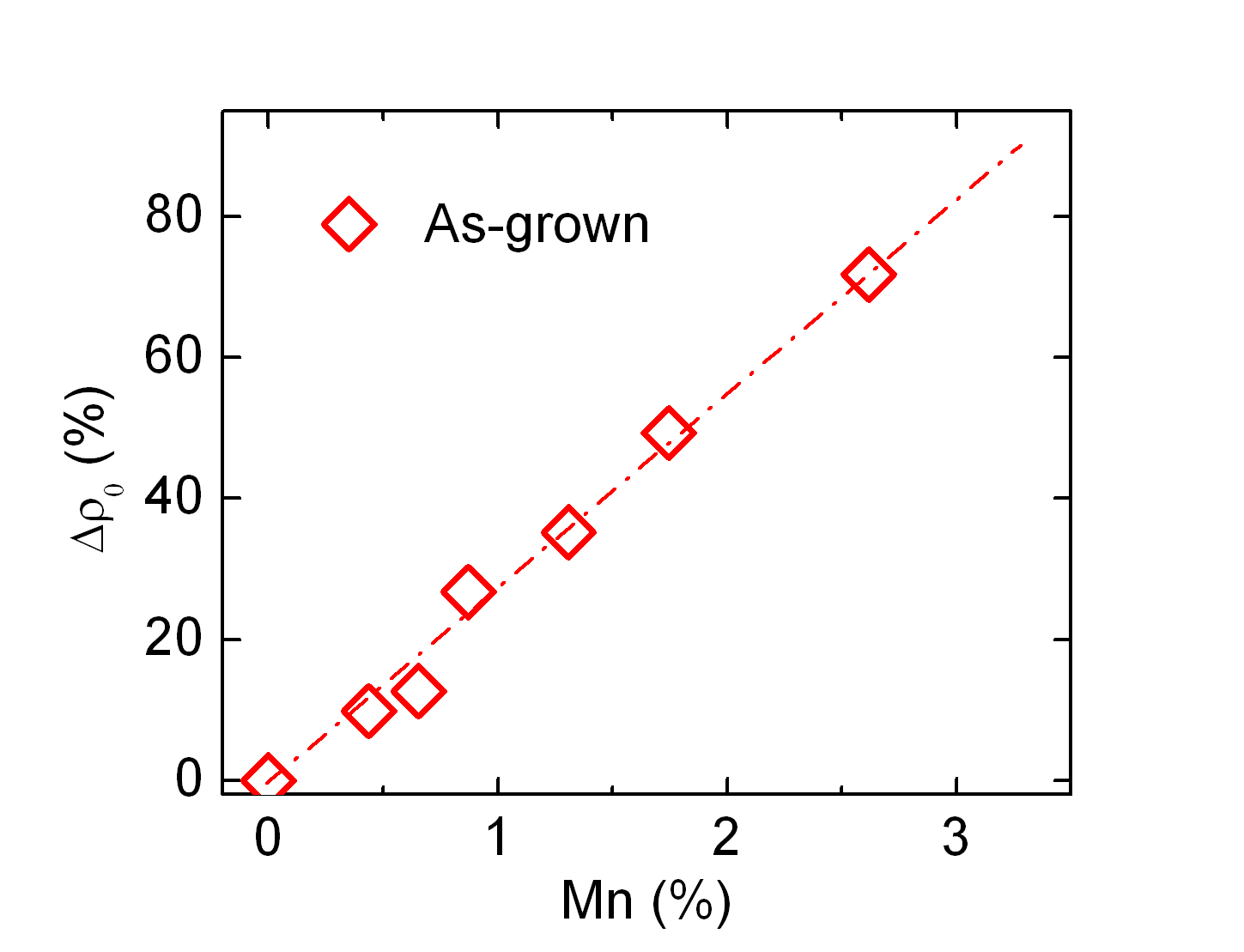}
\caption{{$\bigtriangleup\rho_0 (\%)$} plotted as a function of Mn concentration for the superconducting as-grown crystals (see text for details).}
\label{rho0VsMn}
\end{figure}

The normalized residual resistivity ratio, $\rho_0^n = R(0)/R(300 K)$ is obtained by linearly extrapolating the $\rho^n(T)$ curve from the normal region just above T$ _c $ to $ T = 0 $. The  value of $ \rho_0^n $ is found to increase systematically from $ 0.35 $ for Mn0.0 to $ 0.53 $ for  Mn1.7 sample (see, table \ref{tableII}). The variation is depicted graphically in Fig. \ref{rho0VsMn}, where \% increase in $ \rho_0^n $ ( $= \bigtriangleup\rho_0^n (\%)$), measured with sample Mn0.0 ($ y = 0 $) as a reference, is plotted against the Mn-concentration ($y$), taken in $(\%)$. The value of $\rho^n(T)$ of the sample Mn0.0 is in good agreement with a value of $\sim$0.4 reported by Saha et al. for their optimally Ni-doped Sr122 single crystalline sample  \cite{Saha2009}. It should be stressed that had the resistivity curves shifted rigidly up due to Mn impurities, the quantity $\bigtriangleup\rho_0^n$ would have been zero for all values of temperature; its non-zero value, therefore, signify that the doped Mn impurities cannot be treated as point-like defects -a conclusion which is not surprising but nevertheless important enough to mention.   

The absolute value of the residual resistivity ($\rho_0$) of the superconducting samples is also given in table \ref{tableII}. A residual resistivity value of 165 $\mu \Omega cm$ for sample Mn0.0 compares favorably to the value $\sim$200 $\mu \Omega cm$ reported previously for an optimally Ni-doped Sr122 (ref. \citenum{Saha2009}). With increasing Mn concentration $\rho_0$ increases for the superconducting samples. However, for samples Mn1.3 and Mn1.7 it shows a decrease, which is probably an experimental artifact since the corresponding $\rho_0^n$ shows only a monotonic increase. A significant error in reducing the measured resistance to resistivity can arise if the width of the voltage contacts made using a silver epoxy is larger than  the separation between the probes.  

To get an estimate of T$_c$ suppression rate in terms of change in the residual resistivity, we considered the variation of $\rho_0$ from $165$ for sample Mn0.0 (T$ _c^{zero} $ = 12. 5 K) to $625 $ $\mu \Omega cm$ for sample Mn0.9 (T$ _c^{zero} $ $ \sim$ 5.5 K). Using this data we get a T$ _c $ suppression rate in the range of 15 mK/$\mu \Omega cm$. It should be mentioned that even if we assume an exaggerated error of 50 \% in our $\rho_0$ values, the T$_c$-suppression rate remains smaller than 30 mK/$\mu \Omega cm$, and this rate also falls in the same range if one considers T$_c^{mid}$ rather than T$_c^{zero}$ in our calculation. Values of the same order of magnitude were previously reported for Mn impurities in Ba$ _{0.5}$K$ _{0.5}$Fe$_{2-2x}$M$_{2x}$As$_2$ (in Ref. \citenum{Li2012} a T$_c$ suppression rate of 66 mk/$\mu \Omega cm$ is reported)). As pointed out previously by Prozorov et al. (Ref. \citenum{Prozorov2014}), these  T$ _c $ suppression rates are too low to be reconciled with the AG theory even after taking into consideration the intraband and interband scattering rates.

\begin{table*}[t]
	\caption{\textbf{Residual resistivity and superconducting transition temperature in the samples Sr(Fe, Co, Mn)$_2$As$_2$. Residual resistivity, $\rho_0$ ($\mu \Omega cm$); normalized residual resistivity, ($\rho_0^n$); zero-resistance superconducting onset temperature, T$_c^{zero}$ ($\rho$); superconducting onset temperature (resistivity), T$_c^{onset}(\rho)$; and the superconducting temperature (magnetization), T$_c(\chi)$. In each temperature column the values are in degree Kelvin. $"an"$ in parenthesis indicate annealed crystal}}
	\label{tableII}
	\renewcommand{\arraystretch}{1.3}
	\begin{center}
		\begin{tabular} {p{2cm} p{2cm} p{2cm} p{2cm} p{2cm} c}
			\hline
			Name & \textbf{$\rho_0$} & \textbf{$\rho_0^n$} & \textbf{T$_c^{zero}$ ($\rho$)} & \textbf{T$_c^{onset}(\rho)$} &
		    \textbf{T$_c(\chi)$}\\
			\hline
			Mn0.0 & 165 & 0.35 & 12.5 & 16.5 & 11\\
			%\hline
			Mn0.4 & 170 & 0.39 & 11 & 15 & 9\\
			%\hline
			Mn0.6 & 410 & 0.40 & 8.5 & 11.8 & 7.5\\
			%\hline
			Mn0.9 & 625 & 0.45 & 5.5 & 10.5 & 5 \\
			%\hline
			Mn1.3 & 380 & 0.48 & 2 & 7.5 & $<$ 2 \\
			%\hline
			Mn1.7 & 360 & 0.53 & $<$ 2 & 3 & $< $ 2 \\
			%\hline
			Mn0.0 (an) & 35 & 0.35 & 17 & 19.5 & 16\\
			%\hline
			Mn0.9 (an) & 55 & 0.41 & 10.5 & 14.5 & 10 \\
			%\hline
			Mn1.7 (an) & 160 & 0.49 & 5 & 11 & $<$ 2\\
			\hline
		\end{tabular}
	\end{center}
\end{table*}

\textbf{\begin{figure*}[]
		%\vspace{-0.1cm}
		%\hspace{0.1cm}
		%\begin{center}
		\includegraphics[scale=0.5]{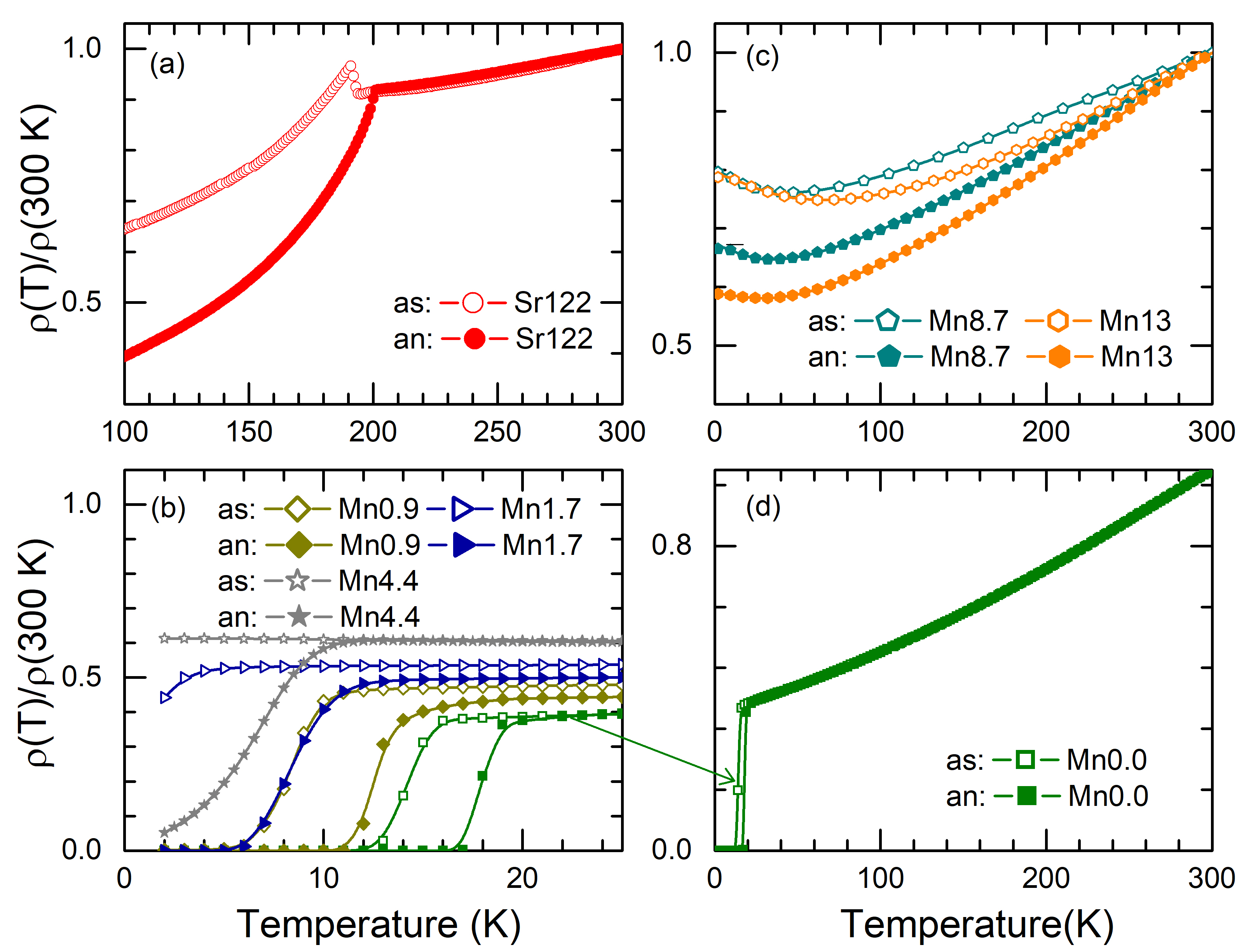}
		%\end{center}
		\caption{Normalized resistivity of as-grown (as), annealed (an) single crystals plotted as a function of temperature for SrFe$_2 $As$_2 $ (panel a), and Sr(Fe$_{0.88-y}$Co$_{0.12}$Mn$_y$)$_2$As$_2$ (panels b and c). Panel (d) shows the normalized resistivity of the asgorwn and annealed Mn0.0 samples over the whole temperature range.}    
		\label{rhoagann}
	\end{figure*}}

\subsection{Effect of low-temperature annealing on T$ _c $ suppression} \label{Effect of gentle annealing on T$ _c $ suppression}

We now investigate the effect of gently annealing the crystals on T$ _c $-suppression rate. Single crystal specimens of samples Sr122, Mn0.0, Mn0.9, Mn1.7, M4.4, Mn8.7 and Mn13 were annealed at a relatively low temperature of 350$^{\circ}$C for 5 days under vacuum in sealed quartz ampules. As shown in the table \ref{tableII}, the residual resistivity values and the superconducting transition temperatures changed significantly upon annealing. We first compare the behavior of sample Sr122 before and after annealing. As shown in panel (a) of Fig. \ref{rhoagann}, the structural/magnetic transition upon annealing increased from 192 K to 200 K. The shape of the anomaly associated with this transition also changed, and now resembles the position and shape previously reported for a Sn-flux grown single crystal \cite{Yan2008, Kim2010, Yan2010}. It is well known that the prolonged annealing at low-temperature relieves strain induced defects. The defect concentration can be particularly high for crystals grown using the self-flux technique due to high processing temperatures involved. Typically, in a growth experiment using the self-flux, the crystal growth ends in the temperature range 900 to 950 $^{\circ}$C. At this point, the entire charge is cooled to room-temperature at a relatively faster rate either by turning off the furnace or by removing the ampule for centrifuging. On the other hand, in a growth experiment involving Sn-flux, the low-melting point of Sn allows for the crystal growth experiment to proceeds down to much lower temperatures (anything between 400 to 500 $^{\circ}$C), where the crystal are decanted from the flux either by centrifuging or by flipping the ampule containing the charge upside-down. Due to a lower processing temperature involved, crystals grown using the Sn-flux are expected to have less strain induced defects than the crystals grown using the self-flux technique. 

The effect of annealing treatment on the superconducting samples is depicted in panel (b) of Fig. \ref{rhoagann}. The annealing treatment shifts the superconducting transition to higher temperatures. For instance, T$ _c $ of sample Mn0.0 increases from 12.5 K to almost 17 K after annealing, i.e., an increase of nearly 26 \%. Simultaneously, the resistivity curve shifts rigidly down due to $ \rho_0 $ decreasing from about 165 to 35 $\mu \Omega cm$. That the resistivity curve is rigidly down-shifted (i.e., through subtraction of a temperature independent contribution in $ \rho(T) $ ) is apparent from the fact that $\rho^n(T)$ curves for the as-grown and annealed crystals overlap over the whole temperature range as shown in panel (d). From this observation it can be inferred that the crystallographic defects in the as-grown crystals are point-like. Strain in FeSCs has been shown to introduce non-magnetic crystallographic defects analogous to the defects produced by electron irradiation \cite{Kim2015}. Even though the concentration of these defects in our as-grown single crystal is not known, an estimation of the T$ _c $-suppression due to these defects can be made by considering the change in T$ _c $ ($\bigtriangleup T_c$ $ \approx $ 4.5 K) and $ \rho_0 $ ($\bigtriangleup\rho_0$ $ \approx $130 $\mu \Omega cm$ ) solely due to annealing. The quantity $\bigtriangleup T_c$/$\bigtriangleup\rho_0$ in this case turns out to be $ \sim $ 35 mK/$\mu \Omega cm$, which is nearly an order of magnitude smaller than 350 mk/$\mu \Omega cm$ reported by Prozorov et al. (Ref. \citenum{Prozorov2014}) for their electron irradiated samples consisting of point-like defects.       

The effect of annealing treatment on T$_c$'s on samples Mn0.9 and Mn1.7 were also examined. In sample Mn0.9, the value of T$ _c^{zero}  $ increased by almost 5 K, while the residual resistivity decreased from 625 to roughly 55 $\mu \Omega cm$. In sample Mn1.7  (Fig. \ref{rhosc}b), which exhibits a superconducting onset near T = 3 K but no zero-resistance state down to 2 K in the as-grown form, now showed a superconducting onset near T = 11 K and a transition to the zero-resistance state near T = 5 K. The magnetic susceptibility of annealed Mn1.7 also show onset of superconductivity near T = 6 K but the superconducting volume fraction remains marginal ($\sim 4 \%$) down to T = 2 K, and the superconducting transition completely disappeared when measured under a high-field of 10 kOe, indicating filamentary superconductivity due to inhomogeneous Mn distribution (see, supplementary figure 11). Similar effect of annealing is manifested by sample M4.4. In its as-grown state, no signs of superconductivity were found down to T = 2 K; however, after annealing a superconducting onset was revealed near T = 10 K but the zero-resistance state remained out of sight down to 2 K. To check if the high onset temperature is spurious or not, we measured the low-field (20 Oe), in-plane susceptibility which showed no signs of diamagnetic signal down to T = 2 K, indicating that the resistive drop is indeed spurious arising from tiny, disconnected puddles of smaller Mn-concentration. These puddles undergo superconducting transition at relatively higher temperatures upon cooling which results in lowering of the resistivity; but since these are disconnected regions, the non-superconducting background of the sample does not allow the zero-resistance state to reach. This is also a reason for the broadening of superconducting transition of the Mn doped crystals, which we alluded to while describing the superconducting transition width in the as-grown crystals. 

\begin{figure}[t]
	\vspace{-0.1cm}
	\hspace{0.1cm}
	\begin{center}
		\includegraphics[scale=0.73]{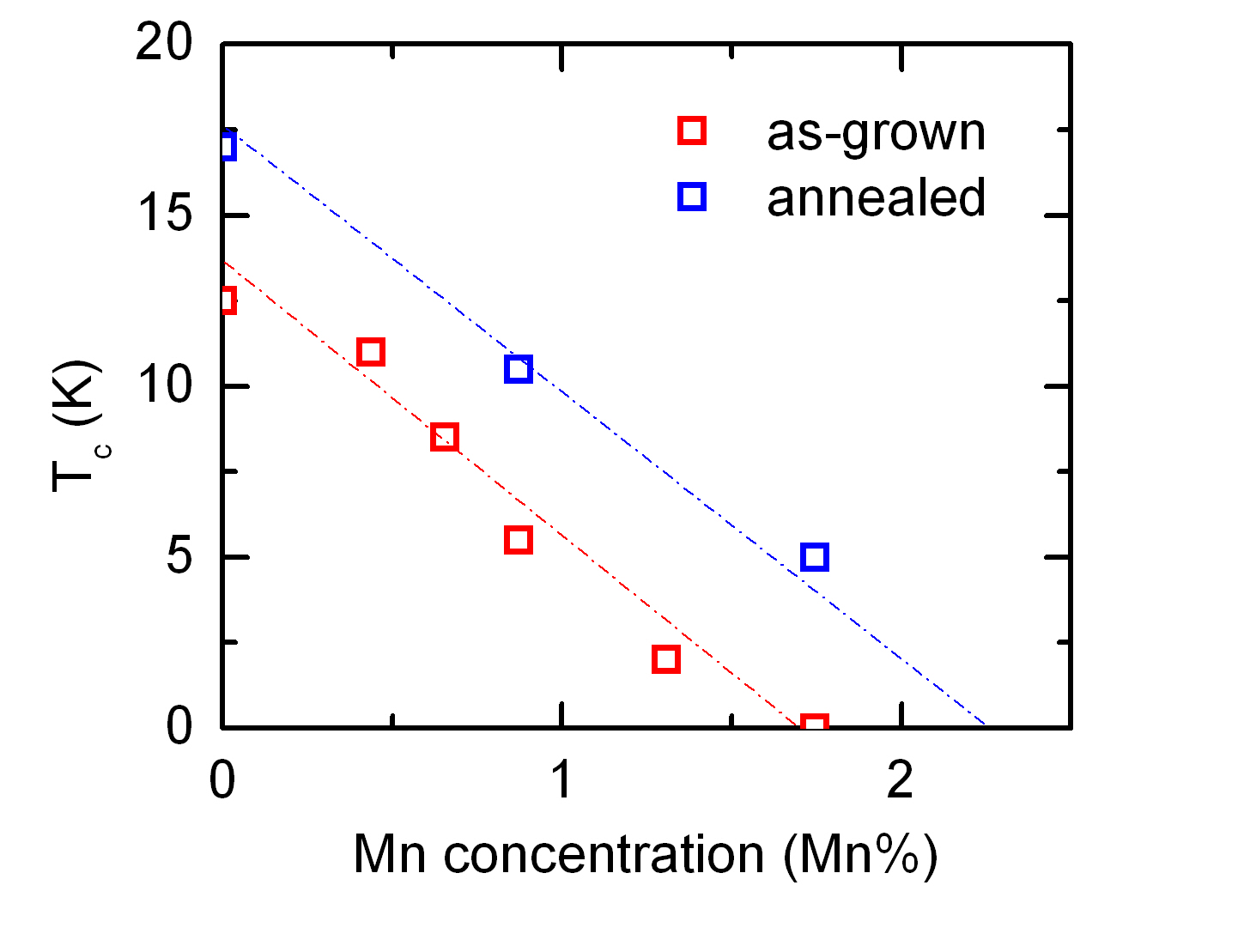}
	\end{center}
	\caption{Plot of T$ _c $ as a function of Mn doping concentration (Mn\%) with respect to Fe for the as-grown and annealed crystals.}
	\label{TCVsMn}
\end{figure}

The occurrence of filamentary or spurious superconducting signal in the annealed samples (Mn1.7 and Mn4.4) may at first sight appear puzzling because the conventional wisdom suggests that annealing treatment homogenizes a sample but here it seems to have an opposite effect. However, the annealing temperature in our experiments is too low to facilitate any kind of atomic diffusion, therefore, the spatial distribution of Mn should not change due to low temperature annealing. What low-temperature annealing does is, it relieves the strain induced crystallographic defects, which lowers the residual resistivity considerably making the spurious superconducting drop in samples with higher Mn-concentration perceptible. What this also suggest is that the low-temperature annealing employed here should not have any bearing on the T$ _c $-suppression effect due to the magnetic scattering. This is in fact the case as is evident from the plot of T$_c$ Vs. Mn-concentration for the as-grown and  annealed crystals shown in Fig. \ref{TCVsMn}. For both set of samples, the T$_c$-suppression rate is almost equal, but the two T$_c$ Vs. Mn-concentration lines are shifted by a constant amount which quantifies the effect of point-like, non-magnetic, crystallographic defects in the as-grown specimens. 

The width of superconducting transitions also decreases as a consequence of annealing the samples. For example, in sample Mn0.0, the ratio $\bigtriangleup T_c/T_c^{mid}$ decreases from 0.27 to 0.14 after annealing. In sample Mn0.9, by way of comparison, this ratio decreases from 0.33 to 0.27. The higher value of $\bigtriangleup T_c/T_c^{mid}$ for the Mn-doped crystal even after annealing treatment is due to inhomogeneous Mn-distribution, which remains unaffected by the low-temperature annealing.  

Finally, we come to the effect of annealing on the non-superconducting samples doped with higher concentrations of Mn. The normalized resistivity of samples Mn8.7 and Mn13 is shown in Fig. \ref{rhoagann}c. For these samples the superconductivity is completely suppressed and their resistivity behavior is rather characterized by the presence of a shallow low-temperature minimum reminiscent of the Kondo effect in the diluted magnetic alloys. At higher Mn-concentration, the Mn-Mn interaction mediated  via the conduction electrons will also come into play. This new energy scale effectively reorganizes the ground state, expelling superconductivity completely. Interestingly, for the non-superconducting samples, the normalized resistivity ($\rho^n$)after annealing differed considerably from the values in the as-grown state, which is different from the as-grown samples where this difference is small or non-existent.   

\section{Discussion} \label{Discussion}

We will now put together the results from the preceding sections to discuss the central question of pairing symmetry in the optimally Co-doped Sr122 superconductor. Our thermopower and Hall data, in line with previous works on Mn substituted FeSCs (see for example: Refs. \citenum{Texier2012,Tucker2012,Rosa2014a, LeBoeuf2014, Gastiasoro2014}), suggest that the \textit{d}-electrons of Mn are localized; and that the charge carrier doping due to Mn substituting Fe in the structure is marginal. Further evidence of the localized nature of Mn \textit{d}-electrons is gathered from the low-temperature magnetization data, which show that the doped Mn-ions are their low-spin state (S = 1/2), which agrees with the previously inferred spin-state using the NMR technique\cite{LeBoeuf2014}. These results are also in agreement with the spin and charge state of Mn impurities doped in the superconducting MgB$_2$ \cite{Rogacki}.  

\begin{figure}[t]
	%	\vspace{-0.1cm}
	%	\hspace{0.1cm}
	\begin{center}
		\includegraphics[scale=0.78]{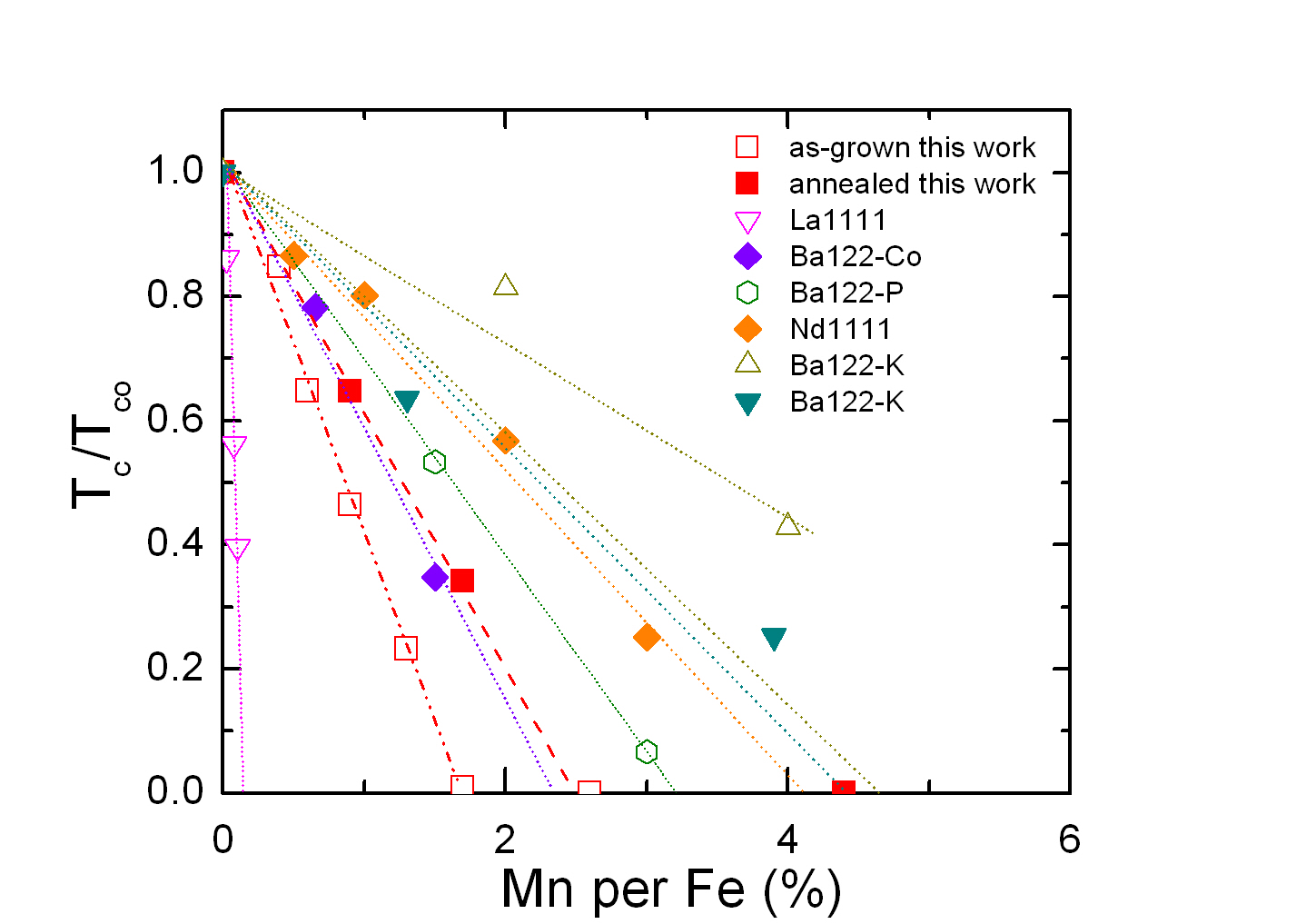}
	\end{center}
	\caption{T$ _c $ suppression rate plotted as a function of Mn concentration expressed as \% per Fe. La1111 (\citenum{Hemmerath2014}), Ba122-Co $\equiv $ Ba(Fe$ _{0.935} $Co$ _{0.065} $)$ 2 $As$ _{2} $ (Ref. \citenum{LeBoeuf2014}), Nd1111(Ref. \citenum{Sato2010}), Ba122-K (up-traingle) $ \equiv  $Ba$ _{0.5} $K$ _{0.5}$Fe$ _2 $As$ _2 $ (Ref. \citenum{Cheng2010}), Ba122-K (down-triangle) $ \equiv $ Ba$ _{0.5}$K$ _{0.5}$Fe$ _2 $As$ _2 $ (Ref. \citenum{Li2012}).}
	\label{FIG.9}
\end{figure}

We shall first discuss whether Mn doping in the present investigation has the same quantitative effect on superconducting T$_c$-suppression as reported previously for other FeSCs or not. For this purpose we plot T$_c$/T$_{c0}$ as a function of Mn-concentration (Mn\%) from the present study along with data collected from literature for several other FeSCs. The results are shown in Fig.\ref{FIG.9}. Here, T$_{c0}$ is the superconducting critical temperature in the absence of Mn impurities. The T$ _c $-suppression rates for the various systems shown in Fig. \ref{FIG.9} are as follows: Ba(Fe$ _{1-x-y} $Mn$ _x $)$ _2 $(As$ _ {1-y}$P$ _y $)$ _2 $  = -10 K/Mn\%  (Ref. \citenum{LeBoeuf2014}),  Ba(Fe$ _ {1-x-y}$Co$ _x $Mn$ _y $)$ _2 $As$ _2 $ =  -12 K/Mn\% (Ref. \citenum{LeBoeuf2014}), Ba$ _{0.5}$K$ _{0.5} $(Fe$ _{1-x} $Mn$ _x $)$ _2 $As$ _2 $ =  -7 K/Mn\% (Ref. \citenum{Li2012}), LaFe$ _{1-x} $Mn$ _x $As(O, F) =  -150 K/Mn \% (Ref. \citenum{Hemmerath2014}), etc. Leaving LaFe$ _{1-x} $Mn$ _x $As(O, F) aside, which we shall come to later, in all other systems the T$_c$-suppression rates are comparable to the rate of approximately -7 K/Mn\% observed in our as-grown and the annealed crystals. The case of LaFe$ _{1-x} $Mn$ _x $As(O, F) is exceptional where a very steep, almost an order of magnitude faster, T$_c$-suppression rate was reported \cite{Sato2010, Hemmerath2014}. This behavior, which has been dubbed as the "poisoning effect", is believed to arise as a consequence of strong electronic correlations unique to this system \cite{Gastiasoro2016}.  

Wang et al. \cite{Wang2013} pointed out that in the multiband superconductors, a more appropriate way of appraising the pair-breaking effect of a certain impurity is by expressing the suppression rates in terms of increase in the residual resistivity: i.e., by how much will the T$_c$ decreases if residual resistivity increases by 1 $\mu \Omega cm$ upon incorporation a certain impurity. This can be understood based on the Abrikosov-Gor'kov formalism, where the superconducting transition temperature in the presence of impurities is governed by the impurity scattering rate \cite{Balatsky2006}. Since the scattering rate itself cannot be directly measured in the experiments, residual resistivity increase ($\Delta \rho_0$) in the presence of impurity doping is a useful quantity to assess the pair-breaking rate due to a certain impurity. However, this criteria can be compared on a quantitative basis provided the impurities are point-like; i.e., the $ \rho(T) $ curve shifts rigidly up upon impurity doping. We show that the normalized resistivity (R(T)/R(300K)) of the as-grown and annealed Mn0.0 crystal overlaps completely in the normal state above the superconducting transition, which suggests that the resistivity curve upon annealing exhibits a rigid down-shift as expected from the point-like defects. We can, therefore, conclude that the as-grown crystal contains point-like crystallographic defects and that these defects can be effectively healed by a low-temperature annealing procedure. This conclusion is in line with the conclusion reached by Kim et al. (Ref. \citenum{Kim2015}) in their recent paper where they showed that the strain induced crystallographic defects in an optimally Co-doped Sr122 are analogous to the non-magnetic, point-defects created due to electron irradiation. We next discuss how this information can be exploited for extracting the gap structure.

In our optimally Co-doped Sr122 crystal, the T$ _c $-suppression due to crystallographic defects was quantified as 35 mK/$\mu \Omega cm$ (see section \ref{Effect of gentle annealing on T$ _c $ suppression}). Intriguingly, this value is nearly an order of magnitude slower than for the electron irradiated Ba(Fe$ _{1-x}$Ru$ _{x}$)$ _{2}$As$ _{2}$ where a T$ _c $-suppression rate of $ \approx $ 350 mK/$\mu \Omega cm$ was obtained. This was shown to be consistent with the s$_{+-}$-wave pairing symmetry. Can the observed slow T$ _c $ suppression rate due to non-magnetic defects found here be reconciled with the s$_{+-}$-wave model is a question which can be answered by investigating  the pair-breaking rate due to the Mn-impurities. 

The actual suppression rate due to Mn is not directly accessible because of unknown amount of residual crystallographic disorder present in the annealed samples; nevertheless, one can try to find a lower bound on it by assuming that in the annealed crystals the only source of pair-breaking is Mn impurities (i.e., the crystallographic defects are completely healed). The absolute value of $ \rho_0 $ for samples Mn0.0, and Mn0.9  is 35 and 55 $\mu \Omega cm$, respectively. Their corresponding T$ _c^zero $'s are: 17 and 10.5 K, which yields an approx. value of 325 mK/$\mu \Omega cm$ as the T$ _c $ suppression rate in the annealed crystals. 

Let us now summarize these findings here: in the as-grown crystals the T$ _c $-suppression rate is around 30 mK/$\mu \Omega cm$, which increases to 325 mK/$\mu \Omega cm$ in the annealed crystals. But our annealing experiment on Mn0.0 sample revealed that the T$ _c $-suppression rate due to crystallographic defects in the as-grown crystals is about 35 mK/$\mu \Omega cm$. From these observations it is obvious that the T$ _c $ suppression in the as-grown crystals is controlled by the crystallographic defects. It is only after these defects have been healed that the \textit{true} T$ _c $ suppression rate due to magnetic impurities can be assessed.    

What can we infer from this analysis? Given a very slow rate of pair-breaking due to crystallographic (essentially non-magnetic) defects found here, together with the finding that the actual T$ _c $ suppression rate due to magnetic impurities is almost an order of magnitude faster, strongly disfavors the $s_{+-}$-wave pairing symmetry. The pair-breaking in $s_{+-}$ state in the presence of magnetic impurities is qualitatively analogous to that in s$_{++}$ state in the presence of non-magnetic impurities \cite{Golubov1997}. The s$ _{++}$ state, on the other hand, is fragile against magnetic impurities but robust against the non-magnetic impurities. This is so because the scattering between the bands of the same sign is equivalent to magnetic pair-breaking in the conventional $ s$-type BCS superconductors, which follows the Abrikoso-Gor'kov (AG) rate \cite{Balatsky2006}. Here, we observe a significant T$ _c $ suppression rate ($\geq$ 325 mK/$\mu \Omega cm$) for the magnetic impurities, and a very slow suppression rate ($\le$ 35 mK/$\mu \Omega cm$) for the non-magnetic crystallographic defects. From these arguments, the s$ _{++}$-wave symmetry seems to be more favorable in the optimally Co-doped SrFe$ _2 $As$ _2 $ superconductor.  
 
These results also show that indeed in FeSCs, the gap symmetry can depend on the specific details of the band structure. For example. the s$ _{+-} $ wave model is applicable to Ba(Fe, Ru)$_2$As$_2$, where Ru substitution at the Fe site is isovalent, and hence no charge carriers are doped. On the other hand, in the optimally electron doped SrFe$ _2 $As$ _2 $,  where the hole-pockets near the Brillouin zone center are substantially shrunk, the gap symmetry seems to be s$ _{++} $.

\section{Summary and conclusions} \label{Summary and conclusions}
The main focus of this manuscript is to investigate the pair-breaking due to magnetic impurities to deduce the gap symmetry. In previous works, T$ _c $ suppression rates for various  transition metal impurities were found to be slow and essentially independent of the magnetic state of the impurity. This result is often cited in the literature as an evidence for the s$ _{+-}$-wave state. We investigated magnetic pair-breaking in optimally electron doped SrFe$ _2 $As$ _2 $ in the presence of various concentrations of magnetic Mn impurities. We show that the pair-breaking rate measured in mK/$\mu \Omega cm$ units agrees fairly well with previous reports on Mn-doping in analogous superconductors.  However, annealing the crystals carefully at low-temperature for several days revealed new information crucial to the determination of pairing symmetry. We show that the strain induced crystallographic defects are a major cause of pair-breaking in the as-grown crystals. We first establishes that these defects are point-like by showing that their effect is to add a temperature independent scattering term that shifts the entire resistivity curves rigidly up. We then estimate the T$ _c $ suppression rate due to these defects and found it to be roughly $\sim$35 mK/$\mu \Omega cm$. However, in the annealed crystals, where the crystallographic defects are healed to a large extent, T$ _c $-suppression rate is found to increase by a factor of 9 or so, which gives a lower-bound on the magnetic pair-breaking rate due to Mn impurities. Interestingly, in both set of crystals (i.e., as-grown and annealed), the T$ _c $-suppression rate measured in terms of doping concentration (\%Mn) remained unchanged, only the lines in the T$ _c $ Vs. \%Mn plot shifted by a constant value due to the point-like crystallographic defects. 

In conclusion, a much slower pair-breaking rate due to non-magnetic defects than expected from the Abrikosov-Gor'kov formalism, together with a pronounced pair-breaking rate due to magnetic impurities disfavors the $s_{+-}$-wave state and favors the $s_{++}$-wave state as the superconducting gap symmetry in the optimally electron doped SrFe$ _2 $As$ _2 $ superconductor. Our work shows that the gap symmetry in FeSCs can indeed vary depending on the details of the band structure. The work also emphasizes that a simple annealing procedure can be very useful in understanding the gap symmetry. In future, it will be interesting to perform similar experiments in annealed Ru doped AFe$_2$As$_2$ superconductors.    

\section*{Acknowledgments}
The authors acknowledge Mr. Nilesh Dumbre and Anit Shetty for their technical assistance. SS is acknowledges financial support by DST-SERB India under grant no. SR/FTP/PS-037/2010.

\section*{References}

\bibliography{Bibliography_Sr122}
\bibliographystyle{unsrt}

\section{Supplemntary Information}
\subsection{Hall effect}
\begin{figure}[]
	% \vspace{-0.1cm}
	% \hspace{2cm}
	\centering
	\includegraphics[height = 7 cm]{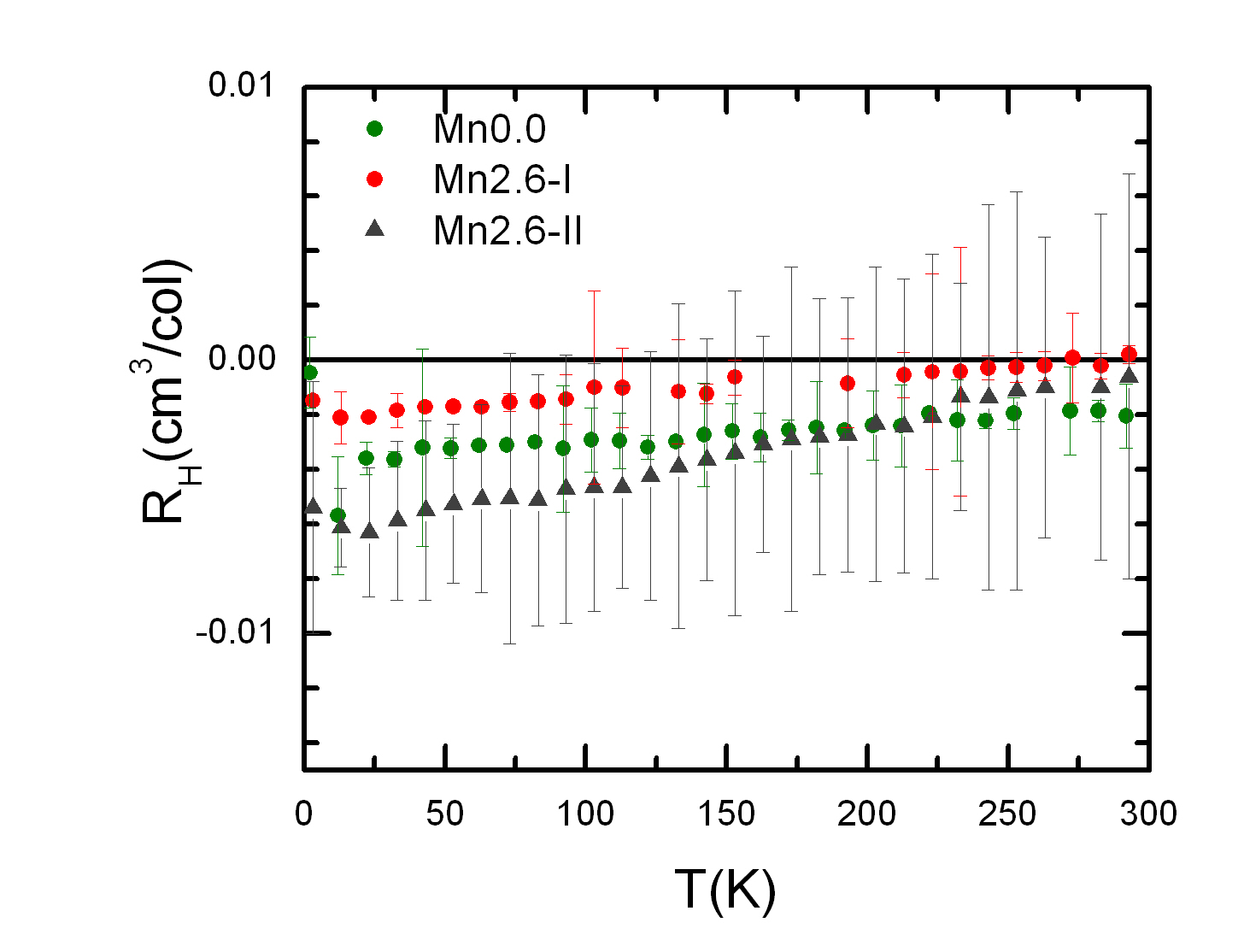}
	\caption{Hall coefficient of Sr(Fe$_{0.88-y}$Co$_{0.12}$Mn$_y$)$_2$As$_2$, y = 0.0, 2.6, single crystals. See text for details.}
	\label{Hall}
\end{figure}

Hall coefficient (R$ _H $) of the crystals was measured using a Physical Properties measurements System (PPMS), Quantum Design, US. Measurements performed under an external magnetic field of 40 kOe applied parallel to the \textit{c}-axis. The longitudinal contribution to the Hall voltage due to contact mismatch was eliminated by reversing the field direction. Temperature variation of R$ _H $ in samples Mn0.0 and Mn2.6 is shown in Fig. \ref{Hall}. The error bar on the R$ _H $ values is obtained from the standard deviation given by the PPMS. Mn2.6-I and Mn2.6-II refers to two different 2.6 \% Mn-doped crystal specimens from the same batch. magnitude of R$ _H $ is small and negative, and is in agreement with the previous results \cite{Kobayashi}. Mn doping has no effect on the value of R$ _H $, which suggest that Mn-doping in SrFe$ _2 $As$ _2 $ does not introduce addition charge carriers in the material.            

\begin{figure}[t]
	%\vspace{-0.1cm}
	%\hspace{2cm}
	\centering
	\includegraphics[height = 5cm]{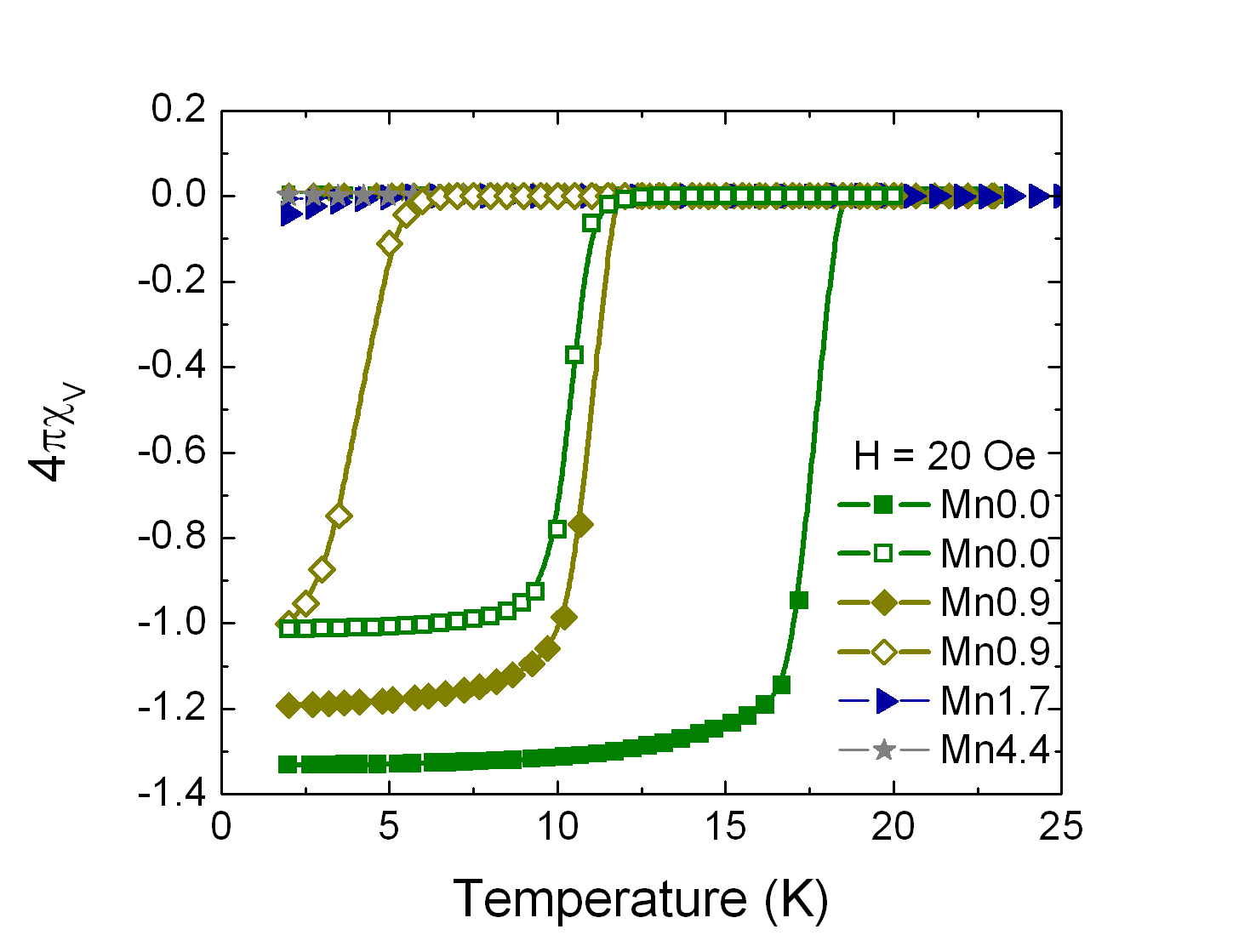}
	\caption{Temperature variation of the volume susceptibility ($ \chi_V $) plotted as $ 4\pi \chi_V $ for Sr(Fe$_{0.88-y}$Co$_{0.12}$Mn$_y$)$_2$As$_2$ (y = 0.0, 0.09, 0.017 and 0.044) annealed single crystals measured under a field (H) of 20 Oe applied parallel to the surface (\textit{ab}-plane). Data for as - grown samples y = 0.0, 0.09, is also shown for comparison.}
	\label{chi-ann}
\end{figure}

\subsection{magnetization}
\begin{figure*}
	%\vspace{-0.1cm}
	%\hspace{2cm}
	\centering
	\includegraphics[height = 10 cm]{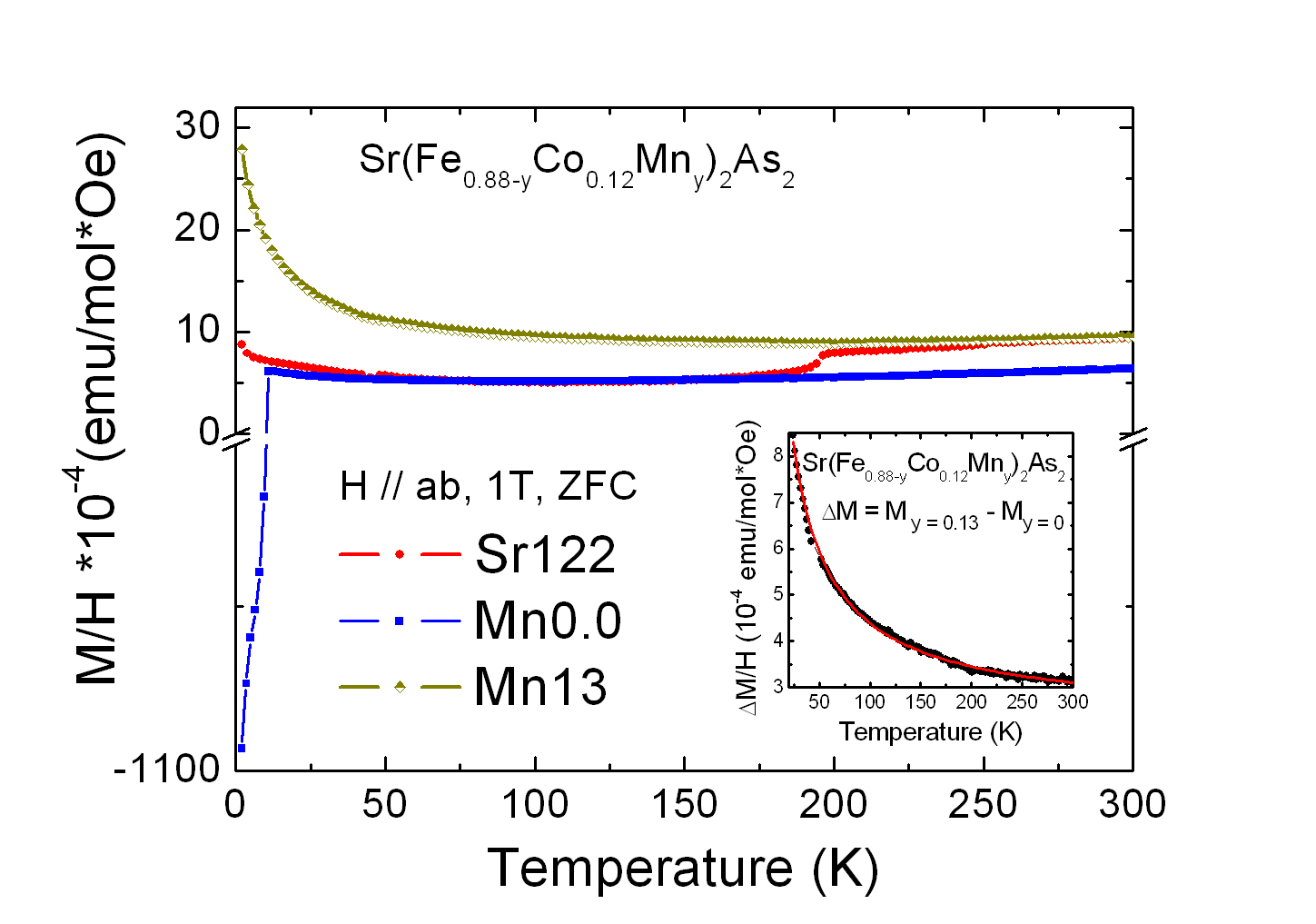}
	\caption{Magnetization data of SrFe$_2$As$_2$ and Sr(Fe$_{0.88-y}$Co$_{0.12}$Mn$_y$)$_2$As$_2$, y = 0.0, 0.13, single crystals measured under an applied magnetic field of 10 kOe, under zero field conditions. The inset shows the Curie - Weiss fit of ($\Delta M$). }
	\label{MbyH}
\end{figure*}

Magnetization of samples Sr122, Mn0.0 and Mn13 is measured under an applied magnetic field of H = 10 kOe (parallel to the \textit{ab}-plane of the crystal) is shown in Fig. \ref{MbyH}. The magnetic behavior of samples with 13 \% Mn doping (Mn13) show a monotonic increase with decreasing temperature without any trace of magnetic or superconducting anomaly down to the lowest measurement temperature of 2 K. The Curie-like increase at low-temperatures in the susceptibility of Sr122 is due to the presence of paramagnetic impurities in the precursory materials. If we assume that the level of these unintended magnetic impurities is approximately equal  in all these samples, then the excess magnetization ($\Delta M$) in Mn13 crystals can be attributed to the doped Mn impurities. 

($\Delta M$) is analyzed using the modified Curie-Weiss law: $\chi = \chi_0 + C/(T - \theta $). $\Delta M$ is obtained by subtracting the measured magnetization of the sample Mn0.0 from that of Mn13 (using Sr122 as a template is avoided due to the magnetic/structural anomaly at $T_0)$. A representative Curie-Weiss fit is shown in the inset of Fig.\ref{MbyH}. The fitting parameters obtained are as follows: $\chi_0 $ = $2.34\times10^{-4}$ $emu/mol\times Oe$, $\theta = -15.4$ K, C = 0.02 $emu/K\times mol\times Oe$. Form the value of Curie constant, the calculated effective moment per Mn came out to be approximately 1.65 $\mu_B/Mn$, which agrees closely with the theoretical effective moment of 1.73 $\mu_B/Mn$ for low-spin state Mn$^{+2}$ in a four-fold coordination. The paramagnetic Curie temperature value is $-15.4 K$, which indicates appreciable antiferromagnetic interactions between the Mn spins. However, as is evident from the M Vs.T plot, no evidence of  magnetic order could be found at low temperatures. The absence of ordering could be due conflicting nature of RKKY-type spin interactions between the randomly distributed Mn spins.

\subsection{Low-field magnetic susceptibility after annealing}
The effect of annealing on the Curie temperature is shown in Fig. \ref{chi-ann}. Date for as-grown samples is also shown for comparison.

\end{document}